\documentclass[aps,pra,twocolumn,groupedaddress]{revtex4-1}
\pdfoutput=1
 \usepackage{amsmath}
 \usepackage{graphicx}   
 \usepackage[english]{babel}

 \usepackage{color}      

\newcommand{\ket}[1]{| #1 \rangle}
\newcommand{\bra}[1]{\langle #1 |}
\newcommand{\ii}{{\mathrm i}}

\begin{document}


\title{Electron pumping in the strong coupling and non-Markovian regime: \\ A reaction coordinate mapping approach}


\author{Sebastian Restrepo}
\email{s.restrepo@tu-berlin.de}
\author{Sina B\"ohling}
\author{Javier Cerrillo}
\author{Gernot Schaller}
\affiliation{Institut f\"ur Theoretische Physik, Technische Universit\"at Berlin, Hardenbergstr. 36, 10623 Berlin, Germany}


\date{\today}

\begin{abstract}
We study electron pumping in the strong coupling and non-Markovian regime. Our model is a single quantum dot with periodically modulated energy and tunnelling amplitudes.  We identify four parameters to control the direction of the current: the driving phase, the coupling strength, the driving frequency and the location of the maxima of the spectral density. In the high-frequency regime, we use a Markovian embedding strategy to map our model to three serial quantum dots weakly coupled to the reservoirs allowing us to use a Floquet master equation. We observe a rectification effect of the pumped charge that is exclusive to the non-Markovian character of our model. In the low-frequency regime, we apply an additional transformation to see our model as three independent transport channels. With the use of full counting statistics, we study charge fluctuations and validate that our model behaves as a single electron source.
\end{abstract}


\maketitle

\section{Introduction}\label{sec:Intro}
Electron pumping is the process of exploiting explicit time dependencies in the Hamiltonian of a system in order to transfer electrons between two different sources.  It has received considerable attention in the past years due to its relevance and potential in quantum nanotechnologies. The primary interest comes from the possibility of creating devices able to manipulate single charges in a precise manner, achieving what is known as a single-electron source \cite{Kouwenhoven1991,Likharev1999,Feve2007}. These devices have significant applications in the field of metrology \cite{Zimmerman_2003,Flowers2004,Keller_2008} which has led to the recent improvement of numerous experimental designs displaying high accuracy single-electron pumping~\cite{Blumenthal2007,Kaestner2008,Yukinori2003,Pekola2007,Giblin2012,Fricke2014,Rossi2014,Yamahata2014,Brun-Picard2016,Kaestner2015,Friederike2015,Friederike2016}. 

Theoretically, transport through time-modulated fermionic systems has been broadly studied \cite{Kohler2005,Camalet2003,Arrachea2006,Cota2005,Sanchez2007,Sanchez2008,Kaiser2006,Gallego2016,Gurvitz2015,Gurvitz2018}.
Particularly, electron pumping is well understood in the low-frequency regime where it has been successfully described with the use of adiabatic (quasi-static) pumping theories \cite{Browwer1998,Aleiner1998,Zhou1999,Makhlin2001,Moskalets2001,Entin-Wohlman2002,Splettsoesser2005,Yuge2012,Riwar2013}.  On the other hand, its study in the nonadiabatic regime remains a challenge. The interest in this regime comes from the fact that the current is proportional to the frequency. Therefore, in order to generate appreciable currents, high accuracy at fast driving is needed. Only a few works have studied single electron pumping in the nonadiabatic regime~\cite{Kaestner2008,Cavaliere2009,Croy2012,Croy2016,Potanina2019}.

The studies referenced before have all relied on the assumption of weak coupling or non-structured (Markovian) reservoirs. In this work, we study a model for electron pumping that goes beyond those two assumptions. More specifically we model an electron pump with a single quantum dot (QD) coupled to two reservoirs with highly peaked spectral densities (SDs) and study transport through the dot in the low and high-frequency regime.  First, in order to capture the strong coupling and non-Markovian effects of our model, we apply a fermionic reaction coordinate (RC) mapping~\cite{Nazir_Gernot2018,Strasberg2018,Schaller2017,Gurvitz2018,Martensen2019} to the reservoirs. The mapping is closely related to the method of time evolving density matrix using orthogonal polynomials algorithm (TEDOPA) \cite{Prior2010,Chin2010,Woods2014,Rosenbach2016} and it has been used as an accurate method for the study of open quantum systems \cite{Martinazzo2011a,Iles-Smith2014,Iles-Smith2016} and thermodynamics \cite{Newman2017,Strasberg2018,Strasberg2016,Strasberg2017a,Schaller2017}. It consists in a redifinition of the boundary between the system and reservoir such that a master equation for the newly redefined system can be used. In the high-frequency regime, to accurately treat the time-dependencies of our model, we make use of Floquet theory for open systems \cite{Grifoni1998,Kohler1997,Gelbwaser-Klimovsky2013,Kosloff2013,Restrepo2016,Restrepo2018} which we can benchmark with the method of time-dependent non-equilibrium Green's functions (TNEGFs)~\cite{Jauho1994,Aoki2014,Kohler2005,Economou2006}. In the low-frequency regime, we apply a second transformation that will allow us to see our model as three independent parallel QDs. Based on the methods of full counting statistics \cite{Esposito2009}, we use an auxiliary equation formulation \cite{Flindt2008,Benito2016,Cerrillo2016} to study the total pumped charge and its fluctuations.

This combination of tools facilitates the recognition of four different means to control the direction of pumping with one of them being a current rectification effect. Notably, we identify a reversal mechanism in the current as the coupling strength between the system and reservoir increases. Also, we observe that the rectification effect occurs exclusively due to the non-Markovian character of our model.

This work is organised as follows: In the next section, we specify the model of study. In Sec.~\ref{sec:Strong Coupling} we characterise our model within the strong coupling and non-Markovian regime, presenting the RC mapping. In Sec.~\ref{sec:CF} we introduce the technique of counting fields and derive equations for the pumped charge and its fluctuations. In Sec.~\ref{sec: high freq} we briefly explain the Floquet master equation used in the long-time limit and present results for the high-frequency regime.  We follow with Sec.~\ref{sec:Series}, where we map our model to three independent QDs and present the results of the low-frequency regime. Finally, in Sec.~\ref{sec:Freq} we discuss the current reversal caused by the frequency of the driving and follow with conclusions.

\begin{figure*}
\begin{center}
  \includegraphics[width=0.99\linewidth]{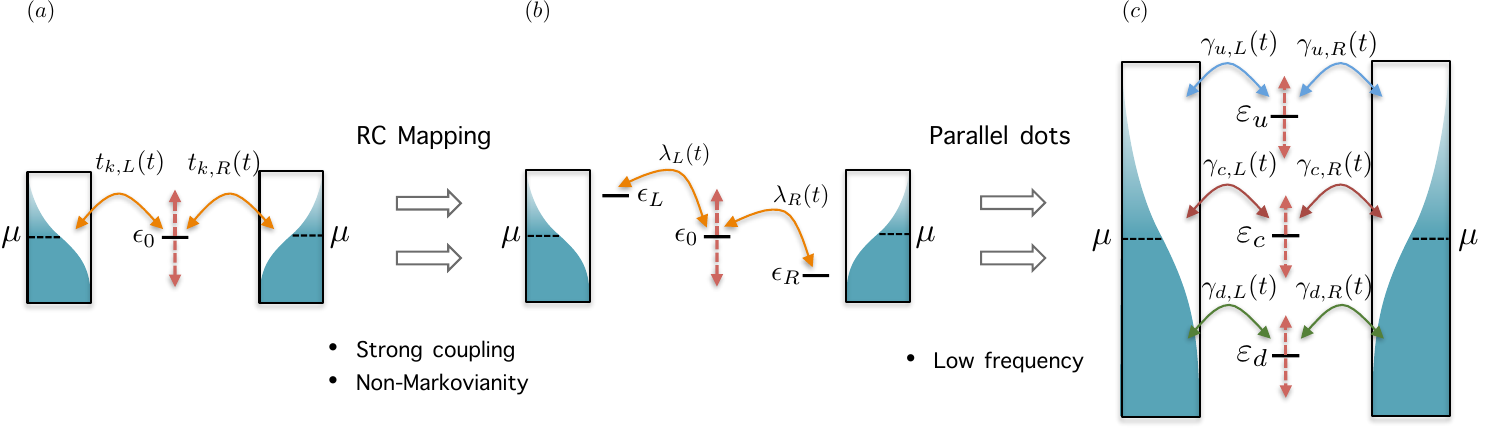} 
\end{center}
\caption{Sketch of the electron pump before and after the two mappings. a: A driven QD is coupled to two reservoirs. b: A driven QD coupled to two reservoirs in the strong coupling and non-Markovian regime is mapped to a triple QD (original QD plus two RCs) now weakly coupled to two residual reservoirs. c: In the low-frequency regime, the triple QD is mapped to three independent parallel transport channels.}
\label{Fig:Maps}
\end{figure*}

\section{Model}\label{sec:Model}
We consider the case of a periodically driven QD coupled to two different baths. The Hamiltonian has the following form
\begin{align} \label{eq:H}
H &= H_S(t) + \sum_{\nu=R,L}  \left[ H_B^{(\nu)} + H_{I}^{(\nu)}(t) \right] \\
 &= \epsilon(t) d^{\dagger}d  +  \sum_{k, \nu} \epsilon_{k,\nu} c_{k,\nu}^{\dagger}c_{k,\nu}, +  \sum_{k, \nu}   \left(  t_{k,\nu}(t) d c_{k,\nu}^{\dagger} +  \rm{h.c.} \right),  \nonumber
  \end{align}
with energies $\epsilon(t)$, $\epsilon_{k,\nu}$, tunnelling amplitudes $t_{k,\nu}(t)$ and where $d$ and $c_{k,\nu}$ are fermionic operators of the system and bath $\nu\in \lbrace L,R\rbrace$, respectively. We assume the Hamiltonian of the system and also its interaction with the bath have a periodic time dependence of the form
\begin{equation} \label{eq:Time-dep}
\begin{aligned}
\epsilon(t) &= \epsilon_0 + a_0 \cos(\Omega  t + \phi), \\ t_{k,R}(t) &= t_{k,R} \left[  1 + a_R \cos(\Omega  t )   \right] , \\
t_{k,L}(t) &= t_{k,L} \left[  1 - a_L \cos(\Omega  t )   \right].  
  \end{aligned}
\end{equation}
where $\Omega$ is the driving frequency. It is related to the period $T$ by $\Omega = 2 \pi / T$. The parameter $\phi$ is the driving phase of the central dot and $a_0$ and $a_{\nu}$ are the driving amplitudes.
We want to study the situation where both baths are at equal chemical potential $\mu$ and inverse temperature $\beta$ and the system is in the Coulomb-blockade regime. Contrary to the case of no driving, where no transport occurs, the time dependencies in the Hamiltonian are able to produce a net matter current between the two reservoirs. We will refer to this transport of electrons at zero bias as \emph{electron pumping}. A sketch of the model is shown in Fig.~\ref{Fig:Maps}a.

To fully characterise the model, the spectral density (SD) of the reservoirs, defined by
\begin{equation}
J_{\nu}(\omega  ) \equiv  2 \pi \sum\limits_k \vert t_{k,\nu} \vert^2 \delta(\omega - \epsilon_{k\nu})  ,
\end{equation}
must be parametrised. The SD contains all the information about how the system and the bath are coupled. A structure-less SD (e.g. flat form) usually allows for a Markovian treatment of the reservoirs due to the fast decay of its associated correlation functions, while a more structured SD (e.g. strongly peaked around a specific frequency) demands a more elaborate treatment. 

Similar models to the one proposed here have been studied before \cite{Croy2012,Croy2016,Potanina2019}. However, all of them were considered in the weak coupling and Markovian regime.  It is our interest to study the case of electron pumping going beyond these approximations with an intuitive method that grants direct access to fluctuations and can potentially also be applied to interacting systems.

\section{Strong coupling and Non-markovian regime }\label{sec:Strong Coupling}
In the following, in order to consider non-Markovian effects, we will consider the case  where the SDs of both baths have a structured form:
\begin{equation}
J_{\nu}(\omega) = \frac{\Gamma_{\nu} \delta_{\nu}^2 }{(\omega-\epsilon_{\nu})^2 + \delta_{\nu}^2},  \label{eq:SD}
\end{equation}
where $\Gamma_{\nu}$ is the coupling strength and $\delta_{\nu}$ the width. The SDs are centered around the energy $\epsilon_{\nu}$. The Markovian limit is obtained taking $\delta_{\nu} \rightarrow \infty$. For finite width $\delta_{\nu}$ we are in the non-Markovian regime~\cite{Zedler2009}. A SD of the form as in equation (\ref{eq:SD}) has been used to study the influence of a measuring lead on a single dot~\cite{Martensen2019,Elattari2000}, molecular wires coupling to electron reservoirs~\cite{Welack2006} and weak coupling approximations in non-Markovian transport~\cite{Zedler2009}. To be able to treat the time dependence in the interaction term adequately and also account for the structured form of the SD, we apply individual fermionic RC  mappings \cite{Nazir_Gernot2018,Strasberg2018,Gurvitz2018} to both baths. The mapping consists in redefining the boundary between the system and the reservoirs such that, at the expense of enlarging the system, we reach a scenario where it is weakly coupled in a Markovian manner to the new \emph{residual reservoir}.

The mapping will only affect the Hamiltonian of the bath and the interaction with the system.  Our model transforms from a single QD coupled to two baths to a triple QD where the right (left) dot is coupled to the right (left) residual bath. The new left and right QDs are referred to as \emph{reaction coordinates} and the new enlarged system is referred to as \emph{supersystem}. A sketch of the RC mapping applied to our model is shown in Fig.~\ref{Fig:Maps}b. After the mapping we have a Hamiltonian of the form
\begin{align} \label{eq:H_mapped}
\tilde{H} = \,& H_{TQD}(t)  + \sum_{k,\nu} E_{k,\nu} C_{k,\nu}^{\dagger} C_{k,\nu} \\ 
&+   \sum_{k,\nu}  \left(  T_{k,\nu}   d_\nu C_{k,\nu}^{\dagger}  + \rm{h.c.}     \right) ,  \nonumber
 \end{align}
where $H_{TQD}(t)$ is the supersystem Hamiltonian of a driven triple quantum dot (TQD), $d_{\nu}$ is a fermionic operator and $C_{k,\nu}$ are fermionic operators of the residual reservoirs. The driven TQD is now our system of interest. Its Hamiltonian is given by 
\begin{align}  \label{eq:HTQD}
H_{TQD}(t) =&  \epsilon(t) d^{\dagger}d + \epsilon_R d_R^{\dagger}d_R +  \epsilon_L d_L^{\dagger}d_L \\
&+ \lambda_R(t) \left( d d_R^{\dagger} +  d_R d^{\dagger} \right)  +  \lambda_L(t)\left(  d d_L^{\dagger}  +  d_L d^{\dagger} \right) , \nonumber
 \end{align}
with
\begin{align}  
  \lambda_R(t) &=\lambda_R \left[  1 + a_R \cos(\Omega  t )   \right], \\
\lambda_L(t) &= \lambda_L \left[  1 - a_L \cos(\Omega  t )   \right].
 \end{align}
All relevant parameters of the mapped Hamiltonian can be obtained in terms of the original SD \cite{Nazir_Gernot2018,Strasberg2018}, see also appendix \ref{App:Mapping} for a brief derivation of the mapping which takes into account a time-dependence in the coupling term. For our particular parametrisation, see equation~(\ref{eq:SD}), we have 
\begin{equation}
\lambda_{\nu} = \sqrt{\Gamma_{\nu} \delta_{\nu} / 2} \quad \text{and } \quad \tilde{J}_{\nu}(\omega) = 2 \delta_{\nu}, \label{eq:SD_mapped}
\end{equation}
where $\tilde{J}_{\nu}(\omega) \equiv  2 \pi \sum\limits_k \vert T_{k,\nu} \vert^2 \delta(\omega - E_{k\nu})$ is the SD of the residual bath. The energies of the two RCs $\epsilon_{\nu}$ are given by the location of the maxima of the original SDs $J_{\nu}(\omega)$.
One of the key features of the RC mapping is that an increase in the interaction strength between the QD and the baths only increases the interaction between the QD and the reaction coordinates. The coupling strength between the RCs and their respective residual reservoirs is unaffected, see equation~(\ref{eq:SD_mapped}).
We will assume that the coupling between the TQD and the residual baths is weak, i.e., $\delta_\nu \beta \ll 1$, such that we can obtain a master equation. The TQD thus follows a time-local master equation that captures the non-Markovian effects of the single QD, our original model.

We point out that even though the RC mapping was applied here to the specific case of a spectral density with a single peak, see equation (\ref{eq:SD}), arbitrary forms may be addressed with it if the use of multiple RCs is considered~\cite{huh2014a,Prior2010,Chin2010,Woods2014,Rosenbach2016}.

\section{Counting Fields}\label{sec:CF}
In order to accurately obtain the current and current fluctuations we follow the full counting statistics formalism \cite{Esposito2009,Flindt2008,Benito2016,Cerrillo2016,Cuetara2015} and introduce a counting field $\xi$ associated with reservoir $\nu$.   Let us define the modified density matrix
\begin{equation}
\rho_{tot}(\xi, t) \equiv   U(\xi,t)\rho_{tot}(0) U^{\dagger}( -\xi,t) , \label{eq:Modif_DM}
\end{equation}
with total (supersystem plus residual reservoirs) initial density matrix $\rho_{tot}(0)$ and modified evolution operator $U(\xi,t) = e^{\frac{\ii}{2}  \xi N_B^{(\nu)}} U(t) \, e^{-\frac{\ii}{2}  \xi N_B^{(\nu)}} $, where $U(t)$ is the evolution operator corresponding to the Hamiltonian of equation~(\ref{eq:H_mapped}) and $N_B^{(\nu)}=\sum_k C_{k,\nu}^{\dagger}C_{k,\nu}$ is the total number operator of reservoir $\nu$. We assume initial factorizing conditions where the reservoirs are described by a thermal state  $\rho_B^{(\nu)} \sim e^{-\beta (H_B^{(\nu)}-\mu N_B^{(\nu)})}$, with inverse temperature $\beta$ and chemical potential $\mu$. Taking the trace over the residual reservoir degrees of freedom, we define the generalized system density matrix
\begin{equation}\label{eq:General_DM}
 \rho(\xi, t) \equiv \text{Tr}_B\left\lbrace \rho_{tot}(\xi, t) \right\rbrace .
\end{equation} 
Note that the total density matrix $\rho_{tot}(t)$ and the reduced density matrix $\rho(t)$ of our system are both recovered by setting the counting field to zero in equations (\ref{eq:Modif_DM}) and (\ref{eq:General_DM}).
We want to obtain the statistics of the electron current. The moment generating function \cite{Esposito2009} associated with the probability $p(\Delta n = n_t-n_0)$ of projectively measuring $N_B^{(\nu)}$ at time $t$ obtaining $n_t$ and at time $0$ obtaining $n_0$ is
\begin{equation}
G(\xi) = \sum_{\Delta n}  e^{\ii \xi \Delta n } p(\Delta n) = \text{Tr}\left\lbrace \rho(\xi, t) \right\rbrace .
\end{equation} 
We define the cumulant generating function of the current
\begin{equation}
 \Phi(\xi) =   \frac{d}{dt} \ln G(\xi) ,
\end{equation} 
so that the cumulants are obtained by simple differentiation 
\begin{equation}
 \frac{d}{dt} \left\langle \left\langle \Delta n^m \right\rangle\right\rangle  =  \left.  \frac{\partial^m}{\partial (\ii \xi)^ m}  \Phi(\xi) \right\vert_{\xi=0} .
\end{equation} 
The generalized density matrix obeys an equation of the form 
\begin{equation}\label{eq:General_rho}
\partial_t \rho(\xi, t)=  \left[ \mathcal{L}(t) + \mathcal{J}(\xi, t) \right]\rho(\xi, t) ,
\end{equation} 
with time dependent superoperator $\mathcal{L}(t)$ and jump superoperator $\mathcal{J}(\xi,t)$, which satisfies $ \mathcal{J}(0,t)= 0$. An auxiliary equation method \cite{Flindt2008,Benito2016,Cerrillo2016} may be used to access high order cumulants and moments directly. For the current $I(t)$ and noise $S(t)$ it follows \cite{Benito2016} 
\begin{equation} \label{eq:Cumulants}
\begin{aligned}
I(t) &\equiv \frac{d}{dt} \left\langle \left\langle \Delta n \right\rangle\right\rangle =  \, \text{Tr} \left\lbrace  \mathcal{J}'(t) \rho(t)   \right\rbrace ,  \\
S(t) &\equiv \frac{d}{dt} \left\langle \left\langle \Delta n^2 \right\rangle\right\rangle =   \text{Tr} \left\lbrace   \mathcal{J}''(t) \rho(t) +2 \mathcal{J}'(t) X(t)     \right\rbrace ,
\end{aligned}
\end{equation} 
where $\mathcal{J}'(t) = \frac{\partial}{\partial (\ii \xi)} \mathcal{J}(\xi,t)\vert_{\xi=0}$, $\mathcal{J}''(t) = \frac{\partial^2}{\partial (\ii \xi)^2} \mathcal{J}(\xi,t)\vert_{\xi=0}  $ and we have introduced the traceless operator $X(t) \equiv    \frac{\partial}{\partial (\ii \xi)} \left[  \rho(\xi,t)/ \text{Tr} \left\lbrace  \rho(\xi,t)\right\rbrace \right]  \vert_{\xi=0} .$ It can be obtained from the auxiliary equation
\begin{equation} \label{eq:Cumulants_X}
\frac{d}{dt} X(t) = \mathcal{J}'(t) \rho(t) -\, I(t) \rho(t) + \mathcal{L}(t)X(t),
\end{equation} 
with initial condition $X(0)=0$. We are interested in the long-time limit $t \rightarrow \infty$. There, we expect the current and noise to acquire the same periodicity as the driving. The total charge $Q$ and its fluctuations $\Delta Q^2$ are then obtained by integrating $I(t)$ and $S(t)$ over a period.

\begin{figure}
  \includegraphics[width=0.999\linewidth]{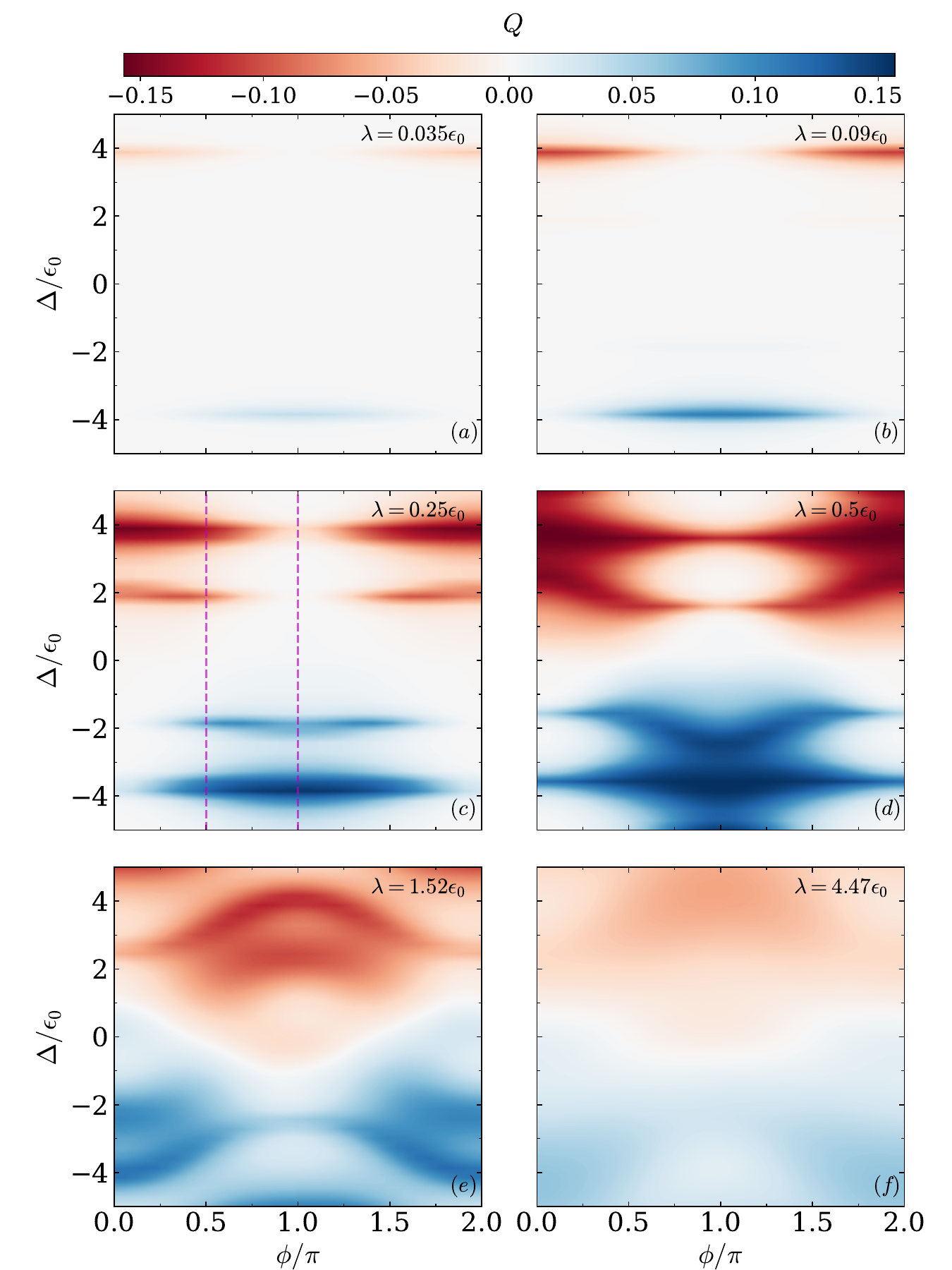}
\caption{Density plots of the total pumped charge during a period $Q$ as a function of the phase $\phi$ and the energy bias $\Delta$ for different values of coupling strength $\lambda$. Positive charge (blue) indicates that electrons are pumped from the left to the right reservoir. Parameters are $\Omega  = 1.9 \, \epsilon_0, \, a_0 = 2.5 \, \epsilon_0, \, \delta = 0.05 \, \epsilon_0, \, \beta \epsilon_0 = 3.3$ and $\mu = \epsilon_0$. Note that the values of coupling strength given correspond to the coupling of the central dot with the left and right dot after the RC mapping. The original coupling to the bath is given by $\Gamma = \lbrace 0.05, 0.324, 2.5, 10, 93, 800 \rbrace \, \epsilon_0$, see equation~(\ref{eq:SD_mapped}). Dashed purple lines correspond to Fig.~\ref{Fig:Density_freq}.}
\label{Fig:Density}
\end{figure}

\section{High frequency - Series TQD}\label{sec: high freq}

To simplify our analysis, we will consider both baths and their coupling strengths $\lambda_{\nu}$ to the central dot to be identical and set the driving amplitudes $a_{\nu}$ equal to one such that we can avoid writing the subscript $\nu = \lbrace R, L\rbrace$ on all parameters appearing in equation~(\ref{eq:SD}), with the exception of the energy $\epsilon_{\nu}$ that corresponds to the energy of the left and right RCs in the mapped model, see equation~(\ref{eq:HTQD}).

To properly take into account the time dependencies in the Hamiltonian of the TQD we use Floquet theory to obtain a master equation. For details the reader is referred to \cite{Grifoni1998,Restrepo2018} or appendix \ref{App:Floquet}. In particular, the method consists in solving the eigenvalue problem for the operator $ H_{TQD}(t) - \ii\partial_t $. Then, we use its eigenstates (Floquet modes) to decompose the interaction between the TQD and the reservoirs and finally, we use this decomposition to derive a master equation for the density operator of the TQD. In the Schr\"{o}dinger picture, the master equation has the form of equation (\ref{eq:General_DM}) for $\xi=0$, where 
$\mathcal{L}(t)$ is a time-periodic superoperator that has the same frequency as the Hamiltonian. In the long-time limit ($t\rightarrow \infty$), we expect  $\rho(t)$ to inherit that same periodicity, such that the master equation may be written as
\begin{equation}\label{eq:Extended}
\ii\, n \, \Omega\, \rho_n = \sum_k \mathcal{L}_k \, \rho_{n-k} \, ,
\end{equation}
where $\rho_n$ and $\mathcal{L}_n$ are Fourier components defined by $\rho(t) = \sum_n e^{\ii  \Omega n t} \rho_n$ and $\mathcal{L}(t) = \sum_n e^{\ii\Omega n  t} \mathcal{L}_n$, respectively. The approach is best suited for high frequencies where convergence in the finite number of Fourier components used is attainable. Studying a regime of low frequency becomes very demanding and inefficient. In the following, the number of Fourier components used was truncated at a level where convergent results were obtained. 

In the standard weak coupling approach,  an additional approximation known as the secular approximation~\cite{Breuer2002} is usually performed. It has the advantage of assuring the master equation has a generator of Lindblad form. It simplifies the study numerically since at steady state one only has to deal with a vector of populations and not the full density matrix. Regarding the RC mapping, applying the secular approximation to the master equation for the supersystem has been shown to fail in obtaining the matter and heat currents accurately~\cite{Nazir_Gernot2018, Strasberg2016, Strasberg2018, Restrepo2018}. We have verified our results using the method of time-dependent non-equilibrium Green's functions (TNEGFs) confirming that the non-secular approach employed here is adequate, see Appendix~\ref{App:Benchmark}. Unlike TNEGFs, the presented method facilitates the calculation of fluctuations (see sec. \ref{sec:CF}) and can potentially also include interactions inside the system.
From now on, all presented results are taken in the long-time limit.


An important feature of our model that would be absent in a Markovian study is the structured form of the SDs, see equation~(\ref{eq:SD}). We parametrize the center of the SDs like $\epsilon_R = \epsilon_0 + \Delta / 2  $ and  $\epsilon_L = \epsilon_0 - \Delta / 2 $. The parameter $\Delta$ will be referred to as energy bias. After the RC mappings, for $\Delta\neq0$, the TQD forms an energy ladder in which the energy of the centre dot and the tunnelling between the centre and left/right dot oscillate as a function of time, see Fig.~\ref{Fig:Maps}b.  This ladder structure adds a new way to control the direction of the matter current. Figure~\ref{Fig:Density} shows density plots of the total pumped charge per period  $Q$ as a function of the driving phase $\phi$ and the energy bias $\Delta$.  Positive charge indicates that electrons are pumped from the left to the right reservoir.

\begin{figure}[t!]
  \includegraphics[width=1.00\linewidth]{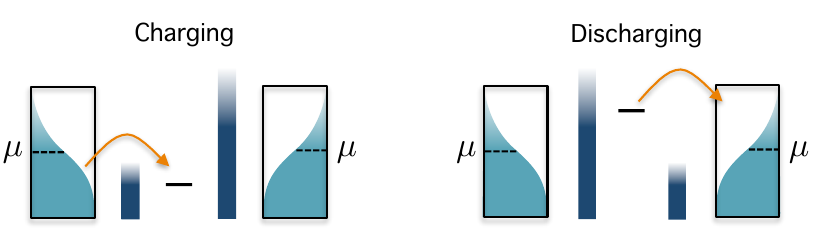} 
\caption{Sketch of the electron pump working in a floodgate manner. The QD gets charged when it is at its lower energy point while one tunnelling barrier is high and the other one is low. When the central dot is at its maximum energy point, the heights of the tunnelling barriers have swapped.}
\label{Fig:Floodgate}
\end{figure}

\subsection{Floodgate behaviour}\label{sec:HF - Floodgate}
Focusing in Fig.~\ref{Fig:Density} on regions where $\Delta \approx 0$, we see that significant pumping only occurs for high values of the coupling strength. Figures~\ref{Fig:Density}e-f shows that the direction of the pumping can then be controlled by the phase of the driving $\phi$. The control of the direction of pumping by the phase obeys the intuitive image of our model in the original picture, see Fig.~\ref{Fig:Maps}a, working in a water lock or floodgate manner where the tunnelling rates can be associated with tunnelling barriers that block or allow the transport of electrons. Figure~\ref{Fig:Floodgate} illustrates this; the central dot is charged by a selected reservoir when it is at its lower energy point while the selected tunnelling barrier is low and the other one is high. When the central dot is at the maximum energy point, the heights of the tunnelling barriers have swapped. The driving phase $\phi$ controls the moment inside the period when the dot is at its maximum or minimum, therefore, controlling the direction of pumping. 

Hereby, we confirm that the ability to control the direction of pumping with $\phi$ extends to the strong coupling and non-Markovian regime at a driving frequency comparable to the energy scales of the system.


\begin{figure}[t!]
  \includegraphics[width=0.999\linewidth]{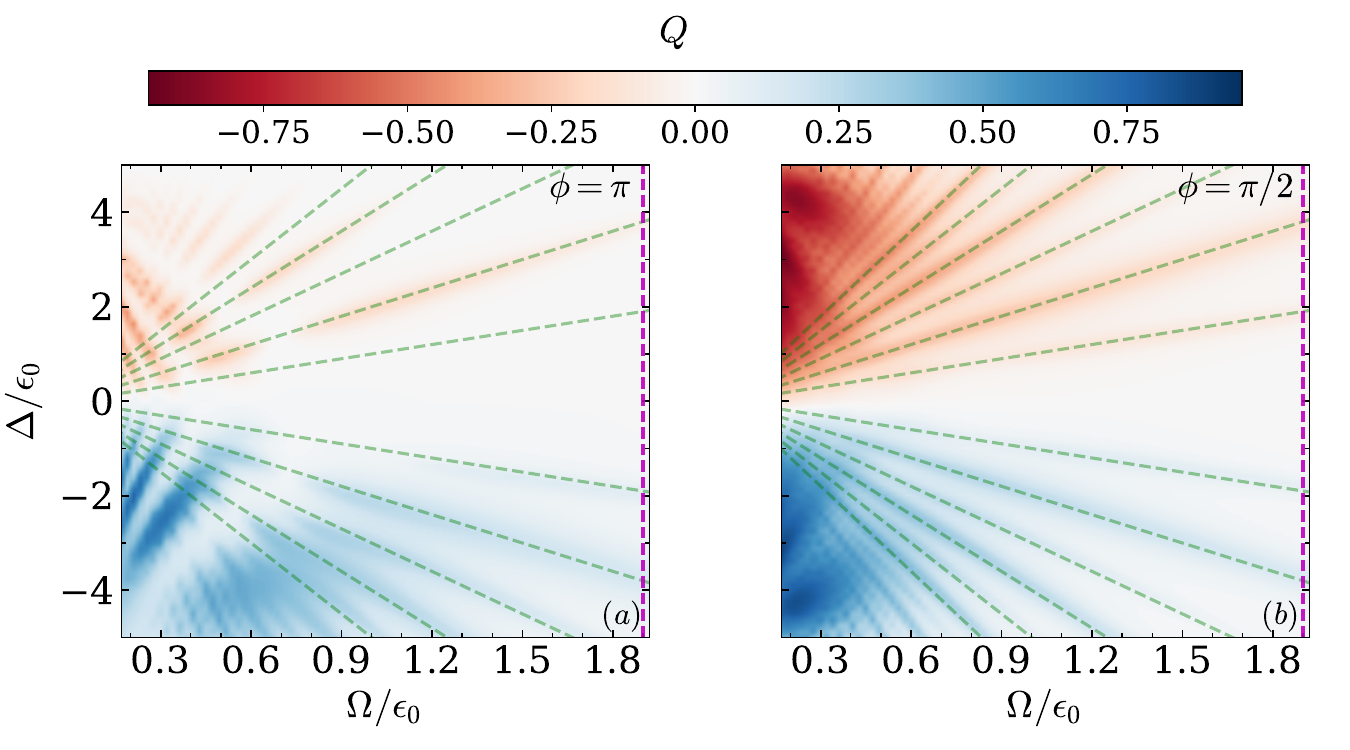} 
\caption{Density plots of the total pumped charge per period $Q$ as a function of the frequency $\Omega$ and the energy bias $\Delta$ for two different phases. The vertical purple lines at $\Omega =1.9 \epsilon_0$ correspond to the vertical purple lines in Fig.~\ref{Fig:Density}c where $\lambda = 0.25 \epsilon_0$. Green dashed lines indicate the resonance condition $\Delta = m \Omega$, with $m$ an integer number.}
\label{Fig:Density_freq}
\end{figure}

\subsection{Current Rectification} \label{sec:HF - Rectification}
Focusing now on the case where $\Delta \neq 0$ we see that there seems to be an optimal coupling strength for achieving maximum transport~\cite{Gelbwaser-Klimovsky2015}. The ladder structure of the TQD induces a preferred direction of pumping regardless of the coupling strength. For $\Delta \gg 0 $ our model acts as a rectifier allowing only the pumping of electrons to the left reservoir and for $\Delta\ll0$ only to the right reservoir.
We also see that pumping seems to occur only at specific resonances. The driving phase $\phi$ no longer controls the direction of the pumping, but it does play a role in the value of the total pumped charge. For a given coupling strength $\lambda$ and energy bias $\Delta$, the driving phase can be critical in achieving pumping. Figures \ref{Fig:Density}a-d show that transport occurs whenever the energy bias $\Delta $ is close to a multiple of the driving frequency $\Omega$, depending on the phase of the driving. Similar resonances have been predicted before in driven triple QDs models where there is a chemical bias, but having a destructive effect on matter transport \cite{Gallego2016}.

\subsection{Single electron source} \label{sec:HF - Single electron source}
For our pump to behave as a single electron source, exactly one electron needs to be pumped every period ($Q=1$) with low charge fluctuations. We assume the electron charge is equal to one. Figure \ref{Fig:Density_freq} shows the total pumped charge per period as a function of the energy bias $\Delta$ and the frequency of the driving $\Omega$. Dashed vertical lines correspond to a driving frequency of $\Omega=1.9\epsilon_0$ previously shown in Fig.~\ref{Fig:Density}c.  We see that as the frequency of the driving decreases, our pump comes closer to this ideal limit. This coincides with previous results obtained for similar models in the weak coupling regime \cite{Croy2012,Croy2016}. 
To study this frequency regime using Floquet theory is exceptionally demanding and inefficient due to a large number of sidebands needed. In sec.~\ref{sec:Series}, we will use an additional transformation to circumvent this problem.

Figure \ref{Fig:Density_freq} also shows (green dashed lines) the resonance condition we saw in Fig.~\ref{Fig:Density}a-d, where maximum pumping was achieved whenever the energy bias $\Delta$ became an integer multiple of the frequency.  We see that the resonance condition loses relevance as the frequency of the driving is decreased, and fewer interferences occur.

\section{Low Frequency - Parallel TQD}\label{sec:Series}
In order to be able to study our model at low frequencies, we perform an additional mapping on the TQD. The new mapping will allow us to regard our model as three independent transport channels, see Appendix~\ref{App:Series}.
The Hamiltonian of the TQD, equation~(\ref{eq:HTQD}),  can be expressed in the following form
\begin{equation} \label{eq:TQD_matrix_eq}
H_{TQD}(t) = \mathbf{d^{\dagger}}\,  \mathbf{H}_{TQD}(t) \,  \mathbf{d},
\end{equation}
where $\mathbf{H}_{TQD}(t)$ is a $3\times 3$ matrix containing all the energies and time-dependent tunneling amplitudes of equation~(\ref{eq:HTQD}) and $\mathbf{d}$ is a 3 dimensional vector containing the annihilation operators $d_L$, $d,$ and $d_R$. Explicit forms of $\mathbf{H}_{TQD}(t)$ and $\mathbf{d}$ are given in Appendix~\ref{App:Series}. After diagonalization we have
\begin{equation} 
H_{TQD}(t)  =    \sum_{i=\left\lbrace u,c,d  \right\rbrace } \varepsilon_i(t) c^{\dagger}_i(t) c_i(t) ,
\end{equation}
where the fermionic operators $c_i(t)$ are obtained by transforming operators $d_L$, $d,$ and $d_R$ with the diagonalization matrix  $\mathbf{T}(t)$. 

The net effect of the unitary transformation $\mathbf{T}(t)$ is a mapping to a parallel QD configuration, where each QD is weakly coupled to the reservoirs, see Fig.~\ref{Fig:Maps}c. However, the time dependence in transformation $\mathbf{T}(t)$ causes the parallel dots to be coupled between each other. 
This becomes clear in the interaction picture with respect to $H_{TQD}(t)$, where the equation of motion of $c_i(t)$ is 
\begin{equation}
\begin{aligned}
\dot{\tilde{c}}_i(t) &=  - \ii\varepsilon_i(t) \tilde{c}_i(t) + \sum_{j,k} \dot{T}_{ij}(t) T^*_{kj}(t) \tilde{c}_k(t),
\end{aligned}
\end{equation}
where a time derivative is denoted by a dot. Whenever $\dot{\mathbf{T}}(t) \approx 0$, which is the case for low driving frequencies, the mapped model can be seen as three independent QDs. To obtain the coupling of the parallel TQD to the residual reservoirs, the transformation $T(t)$ is applied to the residual couplings between the outer dots of the series TQD and the residual reservoirs. 
In the limit of low driving frequency, since the residual coupling is weak, each transport channel can be modeled by a simple rate equation \cite{Kaestner2008,Ohkubo2010,Potanina2019,Croy2012,Croy2016,Kaestner2015} that has the form of equation~(\ref{eq:General_rho}). The pumped charge and its fluctuations are then obtained using equations (\ref{eq:Cumulants}) and (\ref{eq:Cumulants_X}), see Appendix~\ref{App:Series} for the corresponding equations to our model.  The three parallel channels are statistically independent such that the total charge and fluctuations of our model are obtained just by summing the contributions of each channel. 

\subsection{Central dot - Main channel }\label{sec:LF - Central dot}

\begin{figure}[t!]
  \includegraphics[width=0.999\linewidth]{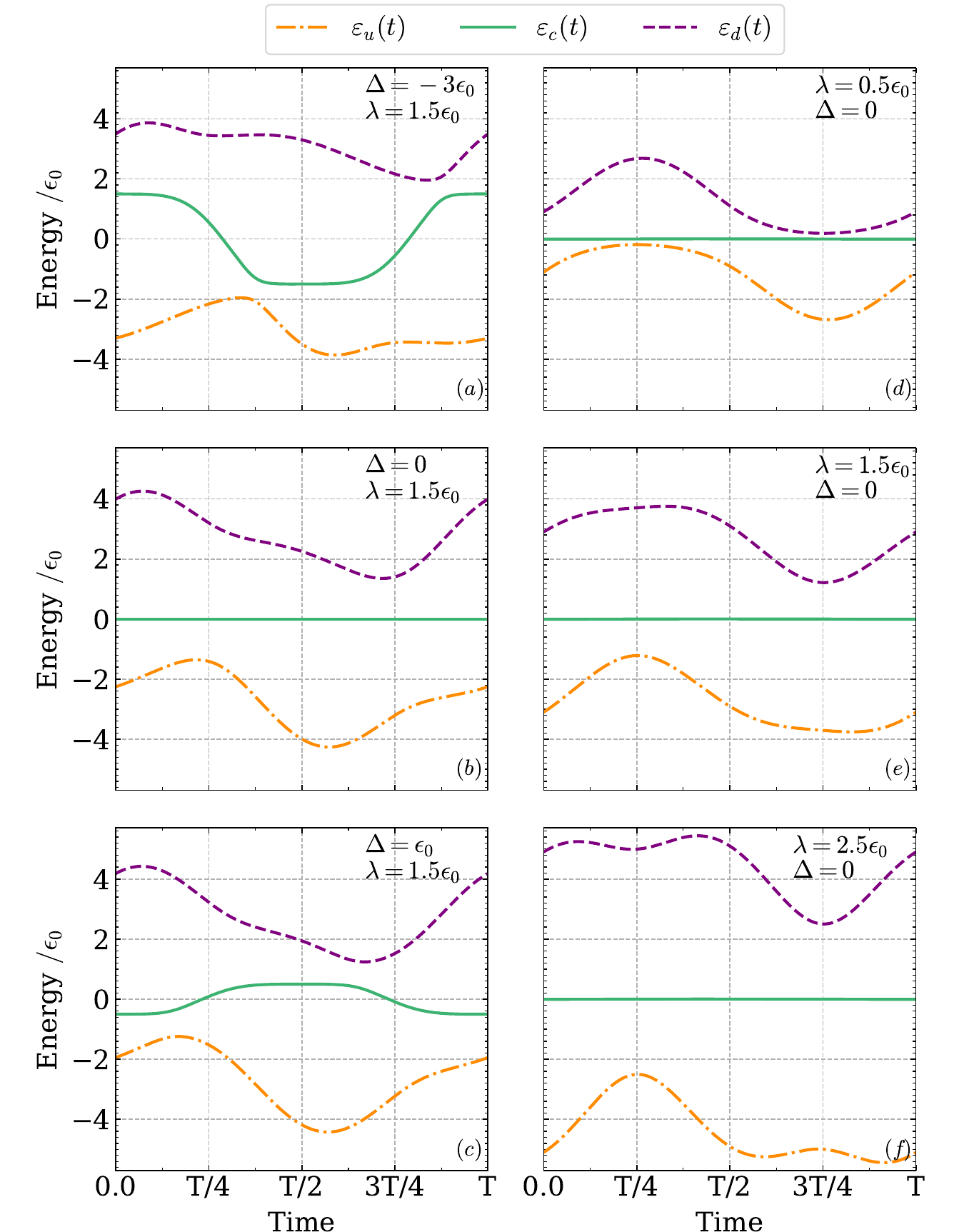} 
\caption{a, b and c: Energy of parallel dots as a function of time for different values of energy bias, $\phi = \pi / 4$ and $\lambda = 1.5 \epsilon_0$. d,e and f: Energy of parallel dots as a function of time for different values of coupling strength, $\phi = \pi / 2$ and $\Delta = 0$.  Other parameters are $\Omega  = 5*10^{-5} \, \epsilon_0, \, a_0 = 2.5 \, \epsilon_0, \, \delta = 0.03 \, \epsilon_0, \, \beta \epsilon_0 = 4,$ and $\mu = \epsilon_0$. In all figures the energy of the central dot is centered around the chemical potential of the reservoir.}
\label{Fig:Energy_Delta}
\end{figure}

Whereas in the high-frequency limit the energy bias $\Delta$ acts as a rectifier (sec.~\ref{sec:HF - Rectification}), in the low-frequency regime it controls the energy levels of the parallel dots. This is shown in the left column of Fig.~\ref{Fig:Energy_Delta}. For $\Delta = 0$, it can be shown that the energy of the central dot has a constant value equal to $\epsilon_0$ such that the central channel does not contribute to transport, see also Fig.~\ref{Fig:Energy_Delta}b, d, e, f.  Figure~\ref{Fig:Charge_Delta}a shows the total pumped charge over a period (blue) and the independent contributions of each channel as a function of the energy bias for a frequency of $\Omega = 5 * 10^{-5}\epsilon_0$. As it was expected, now we are close to pumping one electron per period. We also see that, as the energy bias comes to zero, the pumped charge through the central channel decreases to zero while the contributions from the upper and lower dot increase. The decrease of the central channel is due to its energy becoming constant. For high values of $\Delta$, only the central channel contributes to transport, and the upper and lower dots can be ignored. For transport to occur through the upper and lower channels, their energies $\varepsilon_u(t)$ and $\varepsilon_d(t)$ must come close to the chemical potential of the reservoirs. The left column of Fig.~\ref{Fig:Energy_Delta} shows that as the value of the energy bias moves further away from zero, the energies of the upper and lower dots move away from the chemical potential.


It is important to note that the contribution of the upper and lower dots to transport also has a dependency on the coupling strength. The right column of Fig.~\ref{Fig:Energy_Delta} shows the energy of the parallel dots as a function of time for the specific choice of no energy bias $\Delta=0$ and different values of coupling strength. As the coupling strength increases, the upper and lower dot energies move away from the chemical potential, reducing their contribution to transport. Figure~\ref{Fig:Energy_Delta}d, where $\lambda= 0.5\epsilon_0$, corresponds to the case shown in Fig.~\ref{Fig:Charge_Delta} when we take $\Delta=0$. This proves that for strong coupling and low frequency, a single dot picture applies.

\subsection{Charge fluctuations}\label{sec:LF - Noise}
\begin{figure}[t!]
  \includegraphics[width=1.00\linewidth]{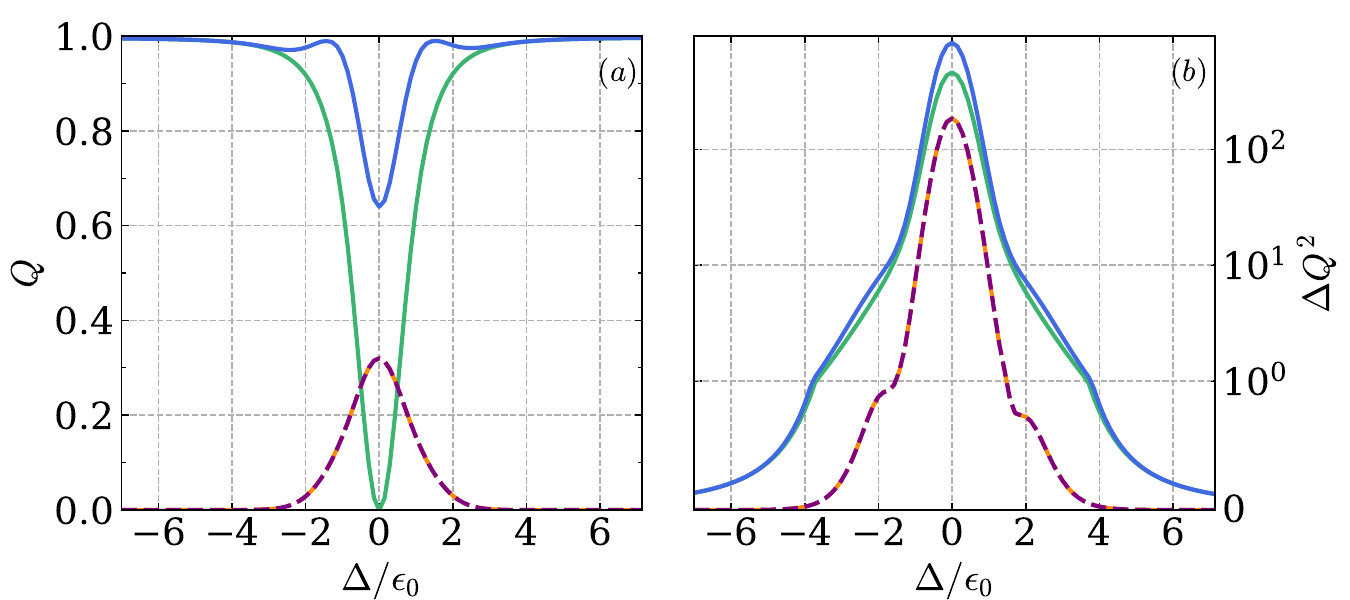}  
\caption{Pumped charge per period (a) and its  fluctuations (b) as a function of the energy bias. Blue lines indicate the total contribution of the three dots. Green lines indicate the contribution of the central dot and orange lines (purple dashed lines) indicate the contribution of the lower (upper) dot. Contributions from the upper and lower dot are indistinguishable. Parameters are $\phi= \pi / 2$ and $\lambda = 0.5\epsilon_0$. Other parameters are the same as in Fig.~\ref{Fig:Energy_Delta}.}
\label{Fig:Charge_Delta}
\end{figure}

Figure~\ref{Fig:Charge_Delta}b shows the charge fluctuations $\Delta Q^2$ as a function of the energy bias.  First, we see that, as it happens for the total pumped charge over a period, the upper (purple dashed line) and lower (orange line) dots behave in an equal manner. As the energy bias approaches a value of zero, their energies come closer to the chemical potential of the reservoir (see Fig.~\ref{Fig:Energy_Delta}) increasing their contribution to transport and their fluctuations. This is a similar scenario to the one encountered in the un-driven model of a single electron transistor where there is a difference in the chemical potentials of the reservoirs. There, the charge and noise increase as the energy of the dot approaches the transport window (defined by the difference in chemical potentials) from the outside.

For the central dot, the behaviour is exactly the opposite. As transport through this channel decreases, the noise goes up. We know that when $\Delta=0$ there is no net transport between the reservoirs. Nevertheless, fluctuations remain high due to the time dependence of the tunnelling rates, now captured all by $\Gamma_{i,\nu}(t)$ (see Appendix~\ref{App:Series}2) since the Fermi function $f(\epsilon_0)$ is constant in time.

Even though the pump is close to transferring one electron per period for all the values of energy bias shown in Figure~\ref{Fig:Charge_Delta}, it can only work as a single electron source for values of energy bias where fluctuations vanish.  It may be tempting to expect the fluctuations of our pump to go down whenever the pump charge per period is close to one. This intuition is only valid in the case of weak coupling, where the transfer of more than one electron per period is unlikely.  In the strong coupling regime, such events are more probable, thus making that scenario possible.

\subsection{Floodgate behaviour}\label{sec:LF - Floodgate}
Figure~\ref{Fig:Density_lambda} shows a density plot of the total pumped charge over a period as a function of the coupling strength and the driving phase. Similar to the case of high frequency, the driving phase remains as a control parameter for the direction of the pumping following a floodgate behaviour as it was explained in sec.~\ref{sec:HF - Floodgate} and illustrated in Fig.~\ref{Fig:Floodgate}. We also see that our pump is now able to transport one electron per period.

\subsection{Coupling strength control - Current reversal } \label{sec:LF - Reversal}
\begin{figure}[t!]
  \includegraphics[width=0.999\linewidth]{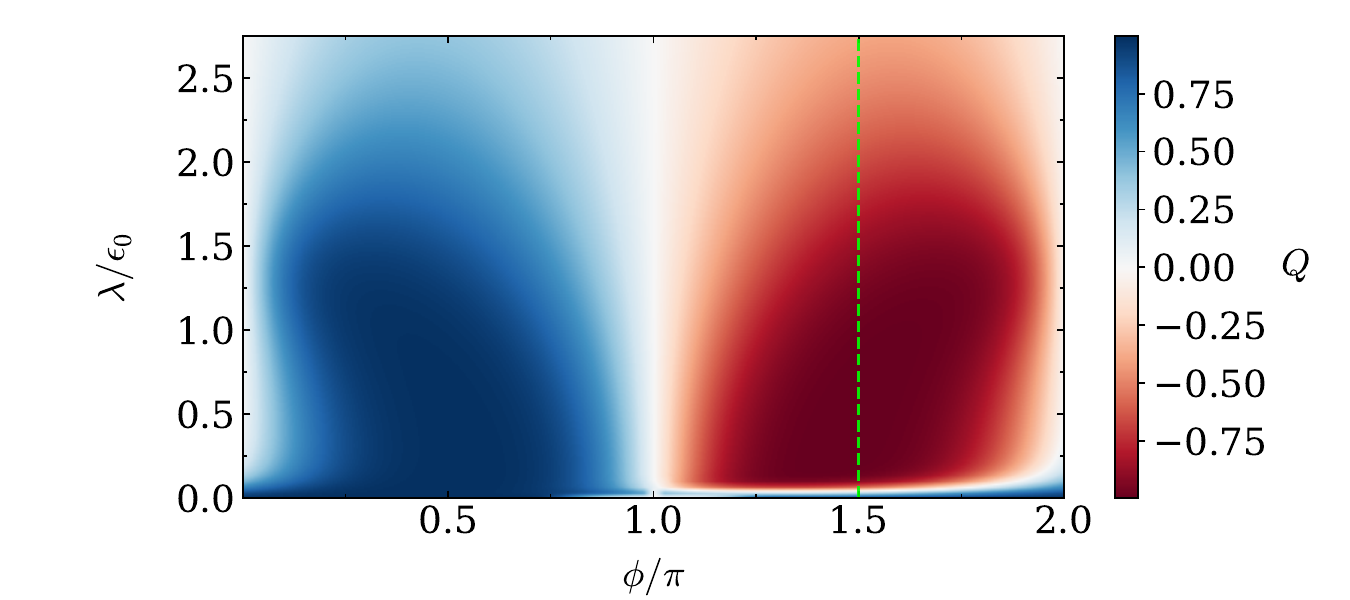} 
\caption{Density plot of the total pumped charge over a period as a function of the driving phase $\phi$ and coupling strength $\lambda$. Note that $\lambda = \sqrt{\Gamma \delta/ 2}$. Parameters are $\Omega  = 5*10^{-5} \, \epsilon_0, \, a_0 = 2.5 \, \epsilon_0, \, \delta = 0.03 \, \epsilon_0, \, \beta \epsilon_0 = 4, \, \Delta = 5 \, \epsilon_0,$ and $\mu = \epsilon_0$. Dashed vertical line corresponds to Fig.~\ref{Fig:Charge_lambda}.}
\label{Fig:Density_lambda}
\end{figure}

Figure \ref{Fig:Density_lambda} also shows a strong dependency between the total pumped charge and the coupling strength. We focus on the case of $\pi < \phi < 2 \pi$, where an increase in the coupling strength reverses the direction of the current. To understand this, we have recognised two transport modes. One that charges/discharges the dot when it is close to the chemical potential and another that occurs when it is at its maximum/minimum of energy. We can cross from one mode to the other by tuning the coupling strength $\lambda$ and thus revert the direction of transport.

\begin{figure}[t!]
  \includegraphics[width=1.00\linewidth]{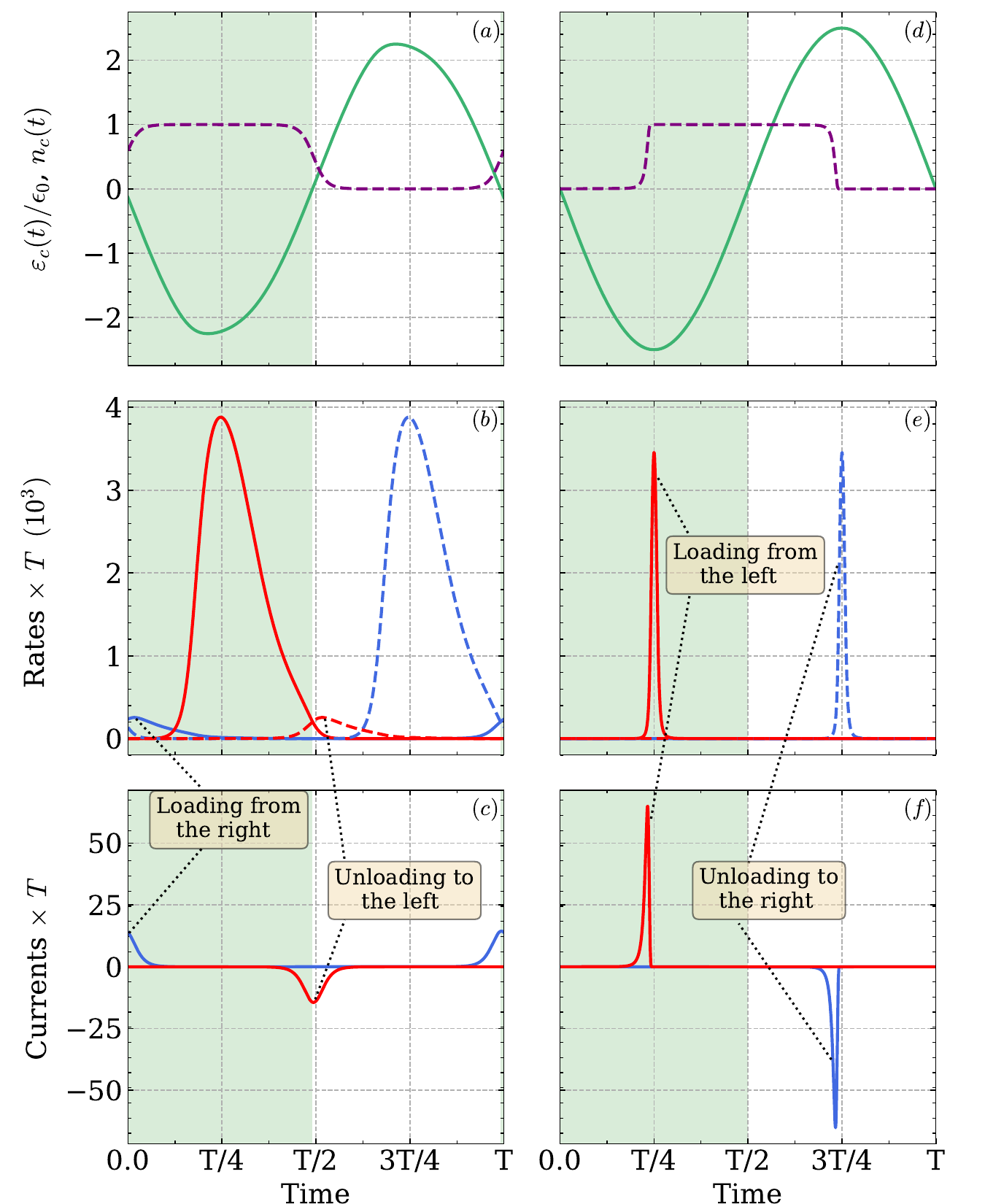} 
\caption{a, d: Energy $\varepsilon_c(t)$ (continuous-green) and occupation $n_c(t)$ (dashed-purple) of the central dot as a function of time. The energy $\varepsilon_c(t)$ is centered around the chemical potential of the reservoirs. b, e: Tunneling rates as a function of time. Continuous lines indicate the rates of tunneling into the dot $f(\varepsilon_c(t)) \Gamma_{c,\nu}(t)$ and dashed line indicate the rates of tunneling from the dot $\left[ 1-f(\varepsilon_c(t))\right] \Gamma_{c,\nu}(t)$. c, f: Matter current of each reservoir as a function of time. Blue (red) lines refer to the right (left) reservoir. Parameters are the same as Fig.~\ref{Fig:Density_lambda} with $\phi = 3 \pi /2$. For the first column we have $\lambda= 0.3 \epsilon_0$ and for the second column $\lambda = 0.003 \epsilon_0$. Green shaded area indicates the time when the central dot is below the chemical potential.}
\label{Fig:Rates}
\end{figure}

We study only the central dot, which captures most of the dynamical features (see sec.~\ref{sec:LF - Central dot}), and look at Fig.~\ref{Fig:Rates}a-c corresponding to $\lambda=0.3 \epsilon_0$ and $\phi = 1.52 \pi$. Figure~\ref{Fig:Rates}a shows the energy of the central dot $\varepsilon_c(t)$ (green) and its occupation (purple) as a function of time for a whole period. We see that the dot gets (un-) occupied whenever the value of the energy crosses the chemical potential of the reservoirs. Once the dot is (un-) occupied, it remains that way until its energy approaches the chemical potential again.
Figure~\ref{Fig:Rates}b shows the rates of an electron tunneling into the dot $f(\varepsilon_i(t))\Gamma_{i,\nu}(t)$ from the right (blue) or left (red) reservoir (continuous lines) and the rates of an electron leaving the dot $\left[ 1-f(\varepsilon_i(t))\right] \Gamma_{i,\nu}(t)$ to the right or left reservoir (dashed lines). It might be tempting to think that the dot is occupied with an electron from the left reservoir since the value of the left rate $f(\varepsilon_c(t))\Gamma_{c,L}(t)$ (continuous red line) is higher than the corresponding value for the right one (continuous blue line) when the dot is at its minimum value of energy. However, the value of the right rate is higher when the energy of the dot approaches from above the chemical potential of the reservoir. The dot is therefore already occupied with an electron from the right bath when the rate of the left bath becomes large, as Fig.~\ref{Fig:Rates}a shows. Figure~\ref{Fig:Rates}c confirms this. It shows a positive right matter current at the beginning of the period coinciding with the moment that the dot gets filled. Conversely, we observe a high rate for the electron tunnelling from the dot to the left reservoir at the moment the energy of the dot becomes higher than the chemical potential of the reservoir, see red dashed line in Fig.~\ref{Fig:Rates}b, giving a net current going from right to left.

The previous case illustrates that looking at the rates whenever the energy approaches the chemical potential defines the direction of the pumping; nonetheless this is not always the case. Both rates (right and left) could be negligible at the moment the energy of the dot crosses the chemical potential, for example when $\lambda = 0.003 \epsilon_0$. Figure~\ref{Fig:Rates}d-f show that for this case of weak coupling the dot gets (un-) occupied from (to) the left (right) reservoir when the energy of the dot is at its minimum (maximum) value, producing a net matter current from left to right that goes to the opposite direction of the case shown in Fig.~\ref{Fig:Rates}a-c.

For a particular choice of driving phase ($\pi < \phi< 2 \pi$), the coupling strength works as a knob to control the direction of the pumping. In the weak coupling case (e.g. Figures \ref{Fig:Rates}d-f) the dot gets occupied from the left reservoir when its energy is at its minimum value but as the coupling strength grows so does the rate of tunnelling from the right reservoir. When the value of this rate becomes non-negligible, the dot starts getting filled from both reservoirs until it gets fully occupied only from the right bath, reversing the direction of pumping. If the coupling strength is further increased, the total pumped charge is reduced until eventually there is no more pumping.

\subsection{Coupling strength control - Fluctuations } \label{sec:LF -  Fluctuations}

\begin{figure}[t!]
  \includegraphics[width=1.00\linewidth]{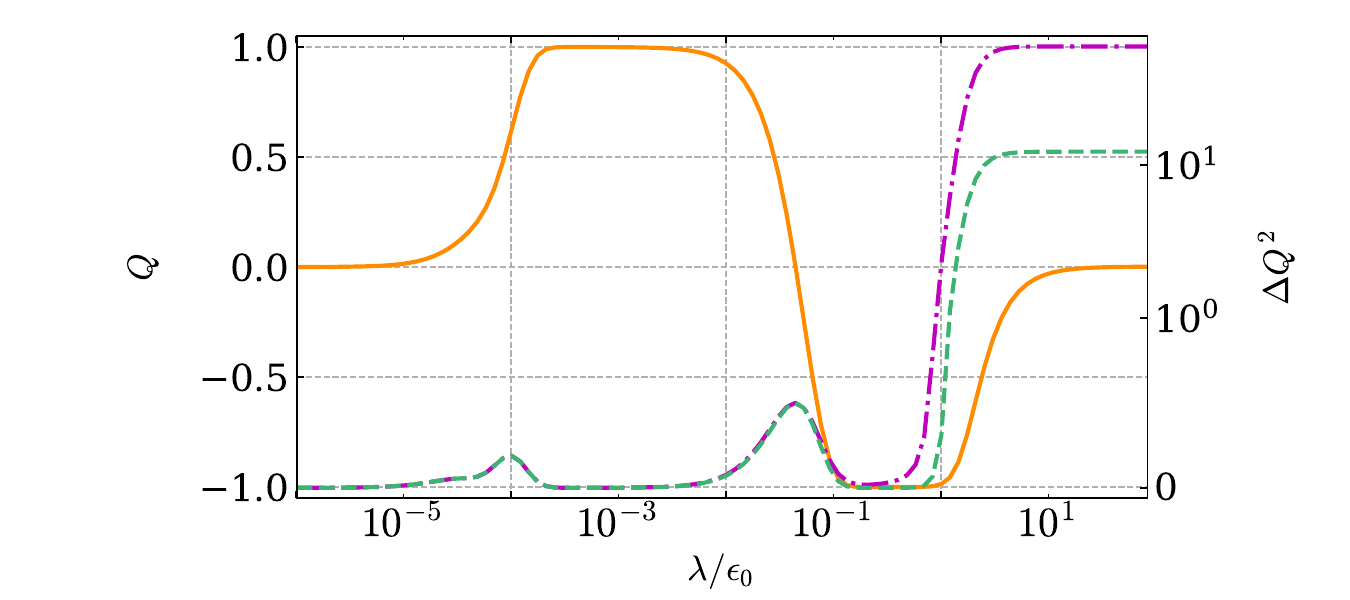} 
\caption{Total pumped charge per period (continuous line) and charge fluctuations (dashed lines) as a function of the coupling strength. Pumped charge corresponds to the dashed vertical line in Fig.~\ref{Fig:Density_lambda} where $\phi = 3 \pi / 2$ and $\beta \epsilon_0 = 4$. Lower charge fluctuations (green) correspond to $\beta \epsilon_0= 20$. The total pumped charge is hardly affected by temperature.}
\label{Fig:Charge_lambda}
\end{figure}

Figure~\ref{Fig:Charge_lambda} shows the total pumped charge over a period as a function of the coupling strength.  It corresponds to the vertical line shown in Fig.~\ref{Fig:Density_lambda} for $\phi = 3 \pi/2$. It also shows the charge fluctuations as a function of the coupling strength.  We see that for different coupling strengths our pump can work as a single electron source transporting one electron every period with low fluctuations.  Contrary to the total pumped charge, fluctuations are strongly affected by temperature (green vs. purple), with fluctuations decreasing for lower temperatures.

\begin{figure}[t!]
  \includegraphics[width=1.00\linewidth]{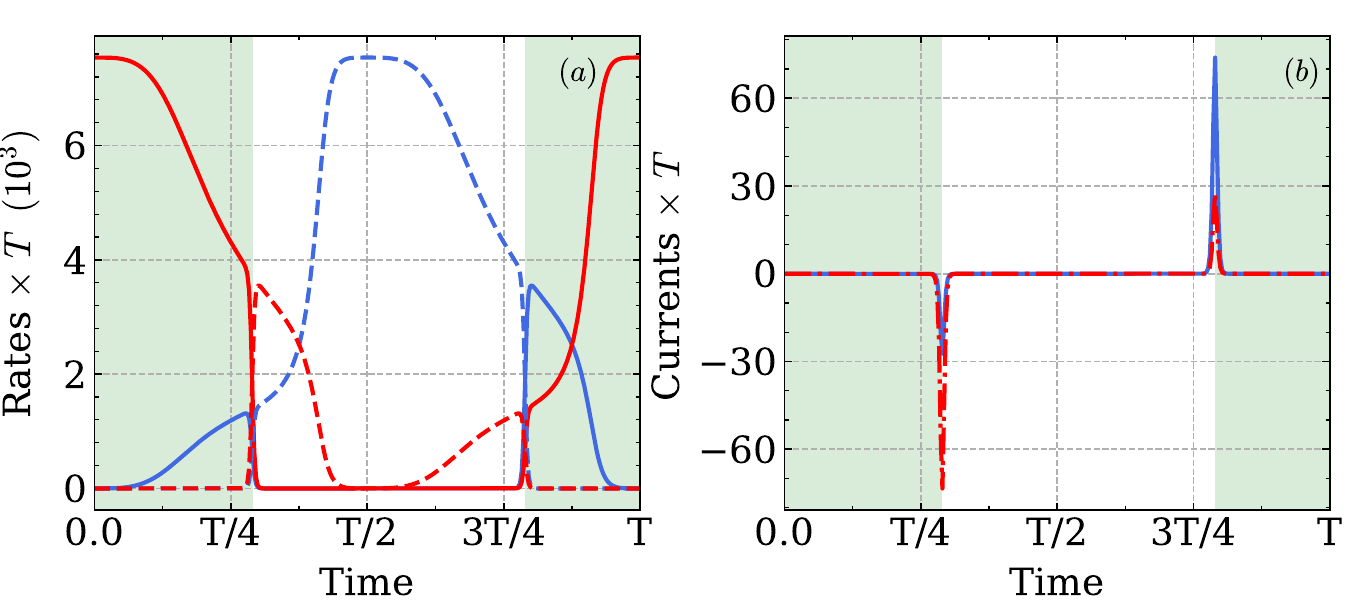} 
\caption{a: Tunneling rates of the central dot as a function of time. Continuous lines indicate the rates of tunneling into the dot $f(\varepsilon_c(t)) \Gamma_{c,\nu}$ and dashed lines indicate the rates of tunneling from the dot $\left[ 1-f(\varepsilon_c(t))\right] \Gamma_{c,\nu}$. b: Matter current of each reservoir as a function of time. Blue (red) lines refer to the right (left) reservoir. Parameters are the same as Fig.~\ref{Fig:Charge_lambda} with $\lambda = 2.5\epsilon_0$ and $\beta \epsilon_0 = 20$. The green shaded area indicates the time when the central dot is below the chemical potential.}
\label{Fig:Rates_noise}
\end{figure}

The high charge fluctuations that occur in the strong coupling regime (Fig.~\ref{Fig:Charge_lambda}) are associated with the processes (charging and discharging the dot) each reservoir is involved in.  Figure~\ref{Fig:Rates_noise} shows the tunnelling rates and currents of each reservoir for a coupling strength of  $\lambda=2.5 \epsilon_0$. In Fig.~\ref{Fig:Rates_noise}a we see that at the moment the energy of the dot crosses the chemical potential (change in background color), the value of the rates from both reservoirs are the same order of magnitude.  This means that the dot gets filled or un-filled with contributions from both reservoirs. Figure~\ref{Fig:Rates_noise}b confirms this by showing high values of current from each reservoir at the same points in time.
Finally, as the coupling strength increases further, the contribution for each reservoir becomes the same such that there is no net charge transferred, but charge fluctuations remain high.

\section{Current reversal - Frequency control}\label{sec:Freq}

In the high-frequency regime, sec. \ref{sec: high freq}, we saw that the direction of pumping can be controlled by the driving phase $\phi$ and that our model worked as a rectifier depending on the sign of the energy bias $\Delta$.  
In the low-frequency regime, sec. \ref{sec:Series}, we saw that the direction of pumping can be controlled by the coupling strength $\lambda$ and the driving phase $\phi$. 

A closer comparative look at figures \ref{Fig:Density_freq}b and \ref{Fig:Density_lambda} suggests that the frequency of the driving $\Omega$ is also a control parameter for the direction of pumping. 
Consider the upper half of Fig.~\ref{Fig:Density_freq}b where $\Delta > 0$ and charge is pumped to the left reservoir. If the frequency is decreased further to the order of $\Omega \approx 10^{-5}\epsilon_0$, the left-hand side of Fig.~\ref{Fig:Density_lambda} where $0<\phi< \pi$ implies that the direction of pumping reverses, now pumping electrons to the right reservoir. 

As mentioned before, looking at the frequency range between the one shown in Fig.~\ref{Fig:Density_freq}b and \ref{Fig:Density_lambda}  is extremely demanding and inefficient with the methods used. Nevertheless, we can assert that a current reversal due to change in the frequency of the driving $\Omega$ exists and extends \cite{Croy2012,Croy2016} to the strong coupling and non-Markovian regime.

\section{Conclusions}\label{sec:Concl}
We have presented a framework to study electron pumping extensively beyond the usual weak coupling and Markovian approximations.  More specifically, we have applied a RC mapping to surmount the lack of separability that exists between system and reservoir in strongly coupled open systems.  For the high-frequency regime, we employed Floquet theory to accurately treat the time dependencies of our model obtaining a non-secular master equation. For the low-frequency case, we applied a second mapping that allowed us to view our strongly coupled model as three independent QDs, for which we investigated the counting statistics.

In both frequency regimes, the model exhibits a floodgate behaviour (Fig.~\ref{Fig:Floodgate}) where the driving phase controlled the direction of pumping. The non-Markovian character of our model gave rise to a rectification of the pumped current in the high-frequency regime (Fig.~\ref{Fig:Density}), where a resonance condition between the driving frequency and the energy bias was observed (Fig.~\ref{Fig:Density_freq}). In the low-frequency regime, although the original time-dependence of our model is harmonic, after the two applied mappings the rates become highly anharmonic (Fig.~\ref{Fig:Rates} and \ref{Fig:Rates_noise}) allowing the identification of suitable control regimes. Our model works as a single electron source (Fig.~\ref{Fig:Charge_Delta} and \ref{Fig:Charge_lambda}), and it was shown that the current could be reversed by changing the coupling strength to the reservoirs.  Finally, we discussed the occurrence of another current reversal now controlled by the frequency of the driving.

Besides the extension of known effects as the floodgate behaviour and frequency current reversal beyond the standard approximations, the observed effects are exclusive to the non-Markovian and strong coupling regime, as the current rectification (from the energy bias) and current reversal (from the coupling strength), provide additional parameters to enhance the performance of electron pumps in the future.

Even though our study centred around a specific kind of driving scheme, the methods used also apply for general time-modulation and in general, provide a formula to study transport in driven systems within the strong coupling and non-Markovian regime in an intuitive manner. 

For the case where interactions are considered, e.g. the QD is occupied by two interacting electrons of a different spin, all the presented methods apply with the caveat that the final mapping in Fig.~\ref{Fig:Maps}c can only diagonalize the quadratic part of the system Hamiltonian. Thus, the three QDs would not enact three independent channels. This type of scenario demands an in-depth study and will be considered in future work.

\acknowledgments{%
 S. R. acknowledges fruitful discussions with C. W. W\"achtler. The authors acknowledge financial support through DFG Grants No. RTG 1558/CRC 910 and BR 1928/9-1.%
 }

\appendix

\section{Fermionic reaction coordinate mapping with time-dependent coupling}\label{App:Mapping}
We present a brief derivation of the fermionic RC mapping with a time-dependent coupling term. For a more detailed description refer to \cite{Nazir_Gernot2018,Strasberg2018}. We consider the case of a general fermionic Hamiltonian of the form
\begin{equation} \label{eq:H_app}
 \begin{aligned}
H &= H_S(t) + H_B + H_{I}(t) \\
 &= H_S(t) +  \sum_{k} \epsilon_{k} c_{k}^{\dagger}c_{k} + \, c \sum_{k} t_{k}(t) c_{k}^{\dagger} - c^{\dagger} \sum_{k} t_{k}^{*}(t)c_{k}  , 
  \end{aligned}
\end{equation}
where $c$ and $c_{k}$ are fermionic operators of the system and reservoir respectively. We assume the Hamiltonian of the system and also its interaction with the bath have an arbitrary time dependence. The mapping we want is such that
\begin{align} \label{eq:H_mapped_app}
H = &H_S(t)  + \lambda^*(t) c d^{\dagger}  + \lambda(t) d c^{\dagger} + \epsilon d^{\dagger}d + \sum_{k} E_{k} d_{k}^{\dagger} d_{k} \nonumber \\
 &+  d \sum_k T_{k} d_{k}^{\dagger} + \sum_k T_{k}^{*} d_{k} \,  d^{\dagger}. 
\end{align}
It is built from a Bogoliubov transformation where fermionic operators $c_k$ are linearly transformed to new fermionic operators $d_k$ as follows
\begin{equation} \label{eq:Bog_transform}
c_k = u_{k,1} d + \sum_{k' > 1} u_{kk'} d_{k'} .
\end{equation}
The transformation $U = (u_{kn})$ is taken to be unitary so that it preserves the canonical commutation relations. From this, comparing equations (\ref{eq:H_app}) and (\ref{eq:H_mapped_app}) and using equation (\ref{eq:Bog_transform}) we have
\begin{equation} 
\vert \lambda(t)\vert ^2 = \sum_k \vert t_k(t)\vert ^2 , \qquad \epsilon = \sum_k \epsilon_k \vert u_{k1} \vert =  \sum_k \epsilon_k \frac{\vert t_{k}(t) \vert ^2}{\vert \lambda(t) \vert^2}. 
\end{equation}
Where we have used the fact that $\sum_k \vert u_{k1} \vert^2  = 1$. To see that the energy of the RC is time independent consider $t_k(t) = f(t) \,t_k  $ with $f(t)$ an arbitrary function of time so that the coupling has then the same time dependence as the tunneling amplitudes $\lambda(t) = f(t) \lambda$. The SD of the bath is defined by $J (\omega  ) \equiv  2 \pi \sum\limits_k \vert t_{k} \vert^2 \delta(\omega - \epsilon_{k})$ such that the RC and its energy are given by
\begin{equation} \label{eq:lambda_epsilon}
 \lambda^2 = \frac{1}{2 \pi} \int J(\omega) d\omega,  \qquad \epsilon = \frac{1}{2 \pi \lambda^2} \int  \omega J(\omega) d\omega.
\end{equation}
To relate the SD of the residual bath, defined by $\tilde{J} (\omega  ) \equiv  2 \pi \sum\limits_k \vert T_{k} \vert^2 \delta(\omega - \epsilon_{k})$, to the SD $J(\omega)$ we look at the Heisenberg equations of motion for the operators in equations (\ref{eq:H_app}) and (\ref{eq:H_mapped_app}). For the original representation, equation (\ref{eq:H_app}), we have
\begin{align} \label{eq:Heisenberg_original}
\dot{c} &= \ii \left[ H_s(t), c  \right] + \ii \sum_k t^*_k \, f(t) c_k   \\
&= \ii S(t)  + \ii \sum_k t^*_k \, f(t) c_k, \nonumber \\
  \dot{c}_k &= - \ii \epsilon_k c_k + \ii t_k \, f(t) c.
\end{align}
A Fourier transformation according to $\int^\infty_{-\infty} \left[ \ldots \right]  e^{\ii z t} dt $ yields
\begin{align} 
z c(z) &=  S(z)  +  \sum_k  \frac{ t^*_k}{2 \pi} \Big( f \ast c_k  \Big)(z) , \\
 z c_k(z) &= - \epsilon_k c_k(z) + \frac{t_k}{2 \pi} \Big(f \ast c \Big)(z),
\end{align}
where the asterisk denotes a convolution $\Big(f \ast h \Big)(z) = \int^\infty_{-\infty} f(z') h(z-z') dz'$. Eliminating the second equation gives
\begin{align} \label{eq:c_original}
z c(z) =&  S(z)  +  \frac{1}{(2 \pi)^2} \int   d\tilde{z}\, f(\tilde{z}) \sum_k \frac{\vert t_k \vert^2}{z-\tilde{z} + \epsilon_k} \\
&\times  \int dz' \, f(z') c(z-\tilde{z} - z'). \nonumber
\end{align}
For the mapped representation, equation (\ref{eq:H_app}), we have
\begin{align} \label{eq:Heisenberg_mapped}
\dot{c} &=  \ii S(t)  + \ii \lambda \, f(t) d, \\
\dot{d} &= - \ii \epsilon d + \ii \sum_k T_k^* d - \ii \lambda f(t) c,  \\
\dot{d}_k &= \ii T_k d - \ii \epsilon_k d_k.
\end{align}
Performing a Fourier transformation we obtain
 \begin{align} 
z c(z) &=  S(z)  +   \frac{ \lambda}{2 \pi} \Big( f \ast d \Big)(z) , \\
  z d(z) &= - \epsilon d(z) + \sum_k T_k d_k(z) - \frac{\lambda}{2 \pi} \Big(f \ast c \Big)(z),\\
   d_k(z) &= T_k d_k(z) - E_k d_k(z ).
 \end{align}
Solving for the system operator $c(z)$ yields the equation
\begin{align} \label{eq:c_mapped}
z c(z) = & S(z)  -  \frac{\lambda^2}{(2 \pi)^2} \int   d\tilde{z}\, 
 \frac{f(\tilde{z})}{z - \tilde{z} + \epsilon -  \sum_k \frac{\vert T_k \vert^2}{z-\tilde{z}  + E_k}  }  \nonumber\\
 &\times  \int dz' \, f(z') c(z-\tilde{z} - z'). 
\end{align}
Comparing equations (\ref{eq:c_original}) and (\ref{eq:c_mapped}) gives the relation
\begin{equation} 
\sum_k \frac{\vert t_k \vert^2}{z-\tilde{z} + \epsilon_k} = 
 \frac{- \lambda^2 }{z - \tilde{z} + \epsilon -  \sum_k \frac{\vert T_k \vert^2}{z-\tilde{z}  + E_k}  } .
\end{equation}
Taking the continuum limit, using the definition of the SDs and evaluating at $z-\tilde{z} = - \omega + \ii \delta $ when $\delta \rightarrow 0^+$ we are able to obtain a relation between the SD of the residual bath with the one from the original representation.
\begin{equation} \label{eq:relation_SD}
\tilde{J}(\omega) = \frac{4 \lambda^2 J(\omega)}{ \left[ \frac{1}{\pi} \mathcal{P} \int_{-\infty}^{\infty} d\omega' \frac{J(\omega')}{\omega'-\omega}\right]^2   +  [ J(\omega)]^2}.
\end{equation}
Equations (\ref{eq:lambda_epsilon}) and (\ref{eq:relation_SD}) allow us to obtain all the relevant parameters of the mapped model in terms of the SD of the original model.

\section{Floquet master equation for fermions with counting field}\label{App:Floquet}
\subsection{Preliminaries}\label{sec:Preliminaries}
We consider a general Hamiltonian of the form
\begin{align} \label{eq:H_sup}
H &= H_S(t) + H_B + H_{I} \\
&= H_S(t) +  \sum_k \epsilon_k c_k^{\dagger}c_k +  \sum_k  \left(  t_k d  c_k^{\dagger} + \rm{h.c.} \right), \nonumber  
\end{align}
where $d$ and $c_k$ are fermionic operators of the system and bath respectively. We assume the Hamiltonian of the system has a periodic time dependence such that we can solve it using Floquet theory.\\

Floquet's theorem establishes that, for a time-periodic Hamiltonian $H_S(t) = \sum_k e^{ \ii k \,\Omega  t} H_k$, with period $T=\frac{2 \pi}{\Omega }$, a solution to Schr\"odinger's equation is given by $ \ket{\psi_r(t)} = e^{- \ii \varepsilon_r t} \ket{r(t)}$, where $\varepsilon_r$ are called quasienergies and $\ket{r(t)}$ Floquet modes (states). The Floquet modes are time periodic and form a complete basis. To find them one solves the eigenvalue problem
\begin{equation}\label{Flo_eigenvalue}
\left(  H_S(t) - \ii \partial_t       \right)  \ket{r(t)} = \varepsilon_r \ket{r(t)}.
\end{equation}
With the Floquet modes at hand, one can find the evolution of any operator. Let us consider an arbitrary operator $S$.
\begin{equation}\label{S_decomp}
S(t) = U^{\dagger}_S(t) S U_S(t) = \sum_{k,l, n }  e^{\ii \Delta_{k,l,n}  t} \, S_{k,l, n}, 
\end{equation}
\noindent where $\Delta_{k,l,n} = \omega_{kl} + n\Omega$ and $S_{k,l, n} = \left[ \int_0^T \frac{dt}{T}  \bra{k(t)} S \, e^{-\ii n \Omega  t} \ket{l(t)}   \right]\, \ket{k}  \bra{l}$. Depending on the form of $\ket{k(t)}$ and $S$, calculating the integral in square brackets might not be trivial.

With decomposition (\ref{S_decomp}) it is now straight forward to obtain a master equation for a driven open quantum system. For a more complete study and review of Floquet theory and driven systems the reader is referred to \cite{Grifoni1998, Eckardt2015}.

Note that
\begin{align}
S^{\dagger}(t) &= \sum_{k,l, n }  e^{-\ii \Delta_{k,l,n}  t} \, (S_{k,l, n} )^{\dagger} \\
& = \sum_{k,l, n }  e^{-\ii \Delta_{k,l,n}  t} \, S_{l,k, -n}^{\dagger} = \sum_{k,l, n }  e^{\ii \Delta_{k,l,n}  t} \, S_{k,l, n}^{\dagger} . \nonumber
 \end{align}

\subsection{Master Equation}\label{sec:ME}
For simplicity and without loss of generality we will assume that there is only one reservoir. The evolution of operator $\rho_{tot}(\xi, t)$ given by equation
\begin{equation}
\partial_t \rho_{tot}(\xi, t) = -\ii \left[   H(\xi,t) \rho_{tot}(\xi,t) - \rho_{tot}(\xi, t)H(-\xi, t) \right] , \label{partial_total_rho}
\end{equation} 
which can be directly obtained by differentiating equation~(\ref{eq:Modif_DM}). Going to the interaction picture and performing the standard Born and Markov approximations \cite{Breuer2002} we obtain  
 \begin{equation}
 \begin{aligned}
 \partial_t \tilde{\rho}(\xi, t) = 
  &- \int^{\infty}_0 ds \text{Tr}_B \lbrace  \tilde{H}_I(\xi,t) \tilde{H}_I(\xi,t-s) \tilde{\rho}(\xi, t) \rho_B   \\
 &- \tilde{H}_I(\xi,t) \tilde{\rho}(\xi, t) \rho_B \tilde{H}_I(-\xi,t-s)  \\
 &-   \tilde{H}_I(\xi,t-s) \tilde{\rho}(\xi, t) \rho_B \tilde{H}_I(-\xi,t)   \\
 &+   \tilde{\rho}(\xi, t) \rho_B \tilde{H}_I(-\xi,t-s) \tilde{H}_I(-\xi,t)  \rbrace. 
 \end{aligned}
 \end{equation}
The interaction picture is defined by $\tilde{A}(t) = U_0^{\dagger}(t)A U_0(t) $, with $U_0(t)$ the evolution operator associated to Hamiltonian $H_0(t)=H_S(t)+H_B$. We define the correlation functions $C_1(\xi,t)= \left\langle \tilde{B}^{\dagger}(\xi,t)B \right\rangle$ and $C_2(\xi,t)= \left\langle \tilde{B}(\xi,t)B^{\dagger} \right\rangle$ with $B = \sum_k t_k^{*} c_k$. Using the fact that $\left\langle \tilde{B}^{\dagger}(\xi,t) \tilde{B}(\xi',t') \right\rangle  = \left\langle \tilde{B}^{\dagger}(\xi- \xi',t-t') B \right\rangle$,  $\left\langle \tilde{B}(\xi,t) \tilde{B}^{\dagger}(\xi',t') \right\rangle  = \left\langle \tilde{B}(\xi- \xi',t-t') B^{\dagger} \right\rangle$ and the form of the interaction Hamiltonian given in equation~(\ref{eq:H_sup}) we have
 \begin{equation}
 \begin{aligned}
 \partial_t \tilde{\rho}(\xi, t) = 
  &- \int^{\infty}_0 ds \, \lbrace \tilde{d}(t)\tilde{d}^{\dagger}(t-s) \tilde{\rho}(\xi, t) C_1(0,s)   \\ & +     \tilde{d}^{\dagger}(t)\tilde{d}(t-s) \tilde{\rho}(\xi, t) C_2(0,s) \\
  &-   \tilde{d}(t) \tilde{\rho}(\xi, t)\tilde{d}^{\dagger}(t-s)  C_2(-2\xi,-s) \\
     &-     \tilde{d}^{\dagger}(t) \tilde{\rho}(\xi, t) \tilde{d}(t-s)  C_1(-2\xi,-s) \\
 &-   \tilde{d}(t-s) \tilde{\rho}(\xi, t)\tilde{d}^{\dagger}(t)  C_2(-2\xi,s)   \\
  &-     \tilde{d}^{\dagger}(t-s) \tilde{\rho}(\xi, t) \tilde{d}(t)  C_1(-2\xi,s) \\ 
  &+ \tilde{\rho}(\xi, t) \tilde{d}(t-s) \tilde{d}^{\dagger}(t)  C_1(0,-s)   \\
  &+   \tilde{\rho}(\xi, t) \tilde{d}^{\dagger}(t-s)  \tilde{d}(t)  C_2(0,-s) \rbrace.  
 \end{aligned}
 \end{equation}
The correlation functions can be easily calculated considering a thermal state for the bath. We obtain
\begin{align}
C_1(\xi,t) &= \int_{-\infty}^{\infty} d\omega \, e^{\ii \omega t} e^{-\ii  \xi} J(\omega) f(\omega), \\
 C_2(\xi,t) &= \int_{-\infty}^{\infty} d\omega \,  e^{-\ii \omega t} e^{\ii \xi}J(\omega) \left[ 1 - f(\omega)\right].
\end{align} 
Here, $f(\epsilon_k) = \left\langle c_k^{\dagger}c_k   \right\rangle = [ e^{\beta(\epsilon_k - \mu )}+1]^{-1}$ denotes the Fermi distribution and we have also introduce the SD of the bath $J(\omega) = \sum_k \vert t_k\vert^2  \delta(\omega - \epsilon_k) $.
Using decomposition (\ref{S_decomp}) only on system operators that have a dependency on $t-s$, performing the integrals over $s$ and $\omega$ with the help of  $\int^{\infty}_0  ds \, e^{\ii \omega s} = \pi \delta(\omega) + \ii \, \mathcal{P}  \, \frac{1}{\omega}$ and disregarding the principal value $\mathcal{P} $ term, one ends up with the equation
 \begin{equation}
 \begin{aligned}
 \partial_t \tilde{\rho}(\xi, t) = &- \sum_{k,l,n}\frac{J(\Delta_{k,l,n})}{2} \, \lbrace \,  f(\Delta_{k,l,n}) [ \, e^{\ii \Delta_{k,l,n} t}  \tilde{d}(t) d^{\dagger}_{k,l,n}\tilde{\rho} \\
 & -  e^{-\ii \Delta_{k,l,n} t}e^{\ii \xi}\tilde{d}^{\dagger}(t) \tilde{\rho}d_{l,k,-n} 
 \\ &-   e^{\ii\Delta_{k,l,n} t}e^{\ii \xi} d_{k,l,n}^{\dagger} \tilde{\rho}\tilde{d}(t) \\
  &+   e^{-\ii \Delta_{k,l,n} t} \tilde{\rho}d_{l,k,-n} \tilde{d}^{\dagger}(t) \, ] \\ 
&+   [1-f(\Delta_{k,l,n}) ] [ \, e^{-\ii \Delta_{k,l,n} t} \tilde{d}^{\dagger}(t) d_{l,k,-n}\tilde{\rho} \\
 &- e^{\ii \Delta_{k,l,n} t}e^{-\ii \xi}  \tilde{d}(t) \tilde{\rho}d^{\dagger}_{k,l,n}\\
 & -   e^{-\ii \Delta_{k,l,n} t}e^{-\ii \xi}  d_{l,k,-n} \tilde{\rho}\tilde{d}^{\dagger}(t) \\
 &+   e^{\ii \Delta_{k,l,n} t} \tilde{\rho}d^{\dagger}_{k,l,n} \tilde{d}(t)\, ] \rbrace.
\end{aligned}
 \end{equation}
To write the equation in the Schr\"odinger picture we need terms of the form $ U_S(t) S_{k,l,n} U^{\dagger}(t) = e^{-\ii \omega_{kl}t} S_{k,l,n}(t)$ with $S_{k,l, n} = \left[ \int_0^T \frac{dt}{T}  \bra{k(t)} S \, e^{-\ii n \Omega  t} \ket{l(t)}   \right]\, \ket{k(t)}  \bra{l(t)}$. It follows then that
 \begin{equation}
 \begin{aligned}
 \partial_t \rho(\xi, t) =&  -\ii [H_S(t), \rho]  \\
  &- \sum_{k,l,n} \frac{J(\Delta_{k,l,n})}{2}\, \lbrace \,  f(\Delta_{k,l,n}) [ \,  d d^{\dagger}_{k,l,n}(t) \rho e^{\ii n \Omega  t} \\
 &-  e^{\ii \xi}d^{\dagger} \rho d_{l,k,-n}(t) e^{-\ii n \Omega  t} -   e^{\ii \xi} d_{k,l,n}^{\dagger}(t) \rho d e^{\ii n \Omega  t} \\
  &+    \rho d_{l,k,-n}(t) d^{\dagger} e^{-\ii n \Omega  t}  ] \\ 
&+   [1-f(\Delta_{k,l,n}) ] [ \, d^{\dagger} d_{l,k,-n}(t) \rho e^{-\ii n \Omega  t}   \\
&-  e^{-\ii \xi}  d \rho d^{\dagger}_{k,l,n}(t) e^{\ii n \Omega  t} -   e^{-\ii \xi}  d_{l,k,-n}(t) \rho d^{\dagger} e^{-\ii n \Omega  t} \\
 &+  \rho d^{\dagger}_{k,l,n}(t) d e^{\ii n \Omega  t}  ] \rbrace.
\end{aligned}
 \end{equation}

The master equation for the system density matrix is obtained by taking $\xi=0$. Defining
 \begin{align}
 \eta_n(t) &\equiv \sum_{k,l}  \frac{J(\Delta_{k,l,n})f(\Delta_{k,l,n})}{2}  d_{l,k,-n}(t), \\
  \theta_n(t) &\equiv \sum_{k,l}  \frac{J(\Delta_{k,l,n})[1-f(\Delta_{k,l,n})]}{2}  d_{l,k,-n}(t), 
 \end{align}
and

\begin{align}
 \tilde{\eta}(t) &\equiv \sum_{k,l}  \frac{\Delta_{k,l,n} J(\Delta_{k,l,n})f(\Delta_{k,l,n})}{2}  d_{l,k,-n}(t), \\
  \tilde{\theta}_n(t) &\equiv \sum_{k,l}  \frac{\Delta_{k,l,n} J(\Delta_{k,l,n})[1-f(\Delta_{k,l,n})]}{2}  d_{l,k,-n}(t), 
  \end{align}

\noindent we have
 \begin{equation}
 \begin{aligned}
 \partial_t \rho(t) =  &-\ii [H_S(t), \rho]   \\
 &+ \sum_{n} \lbrace \,  e^{-\ii n \Omega  t}  \left( [d^{\dagger}, \rho(t)\eta_n(t)]+ [\theta_n(t)\rho(t), d^{\dagger}] \right) \\ 
 &+ e^{\ii n \Omega  t} \left(  [\eta^{\dagger}_n(t) \rho(t), d] 
 + [d, \rho(t) \theta^{\dagger}_n(t)] \right)   \rbrace.
\end{aligned}
 \end{equation}
For the currents we have
 \begin{equation}
 \begin{aligned}
 \partial_t \left\langle N_B \right\rangle  =  &\sum_{n} \text{Tr}  \lbrace \,  e^{-\ii n \Omega  t} \left(  d^{\dagger} \rho(t)\eta_n(t)  - \theta_n(t)\rho(t) d^{\dagger} \right) \\
 &+ e^{\ii n \Omega  t} \left( \eta^{\dagger}_n(t) \rho(t) d   - d \rho(t) \theta^{\dagger}_n(t) \right) \rbrace.
\end{aligned}
 \end{equation}
 \begin{equation}
 \begin{aligned}
 \partial_t \left\langle H_B \right\rangle  = &\sum_{n} \text{Tr} \lbrace \,  e^{-\ii n \Omega  t} \left(  d^{\dagger} \rho(t)\tilde{\eta}(t)  - \tilde{\theta}_n(t)\rho(t) d^{\dagger} \right) \\
 &+ e^{\ii n \Omega  t} \left( \tilde{\eta}^{\dagger}_n(t) \rho(t) d   - d \rho(t) \tilde{\theta}^{\dagger}_n(t) \right) \rbrace.
\end{aligned}
 \end{equation}


\section{Benchmark}\label{App:Benchmark}

\begin{figure}[t!]
  \includegraphics[width=1.00\linewidth]{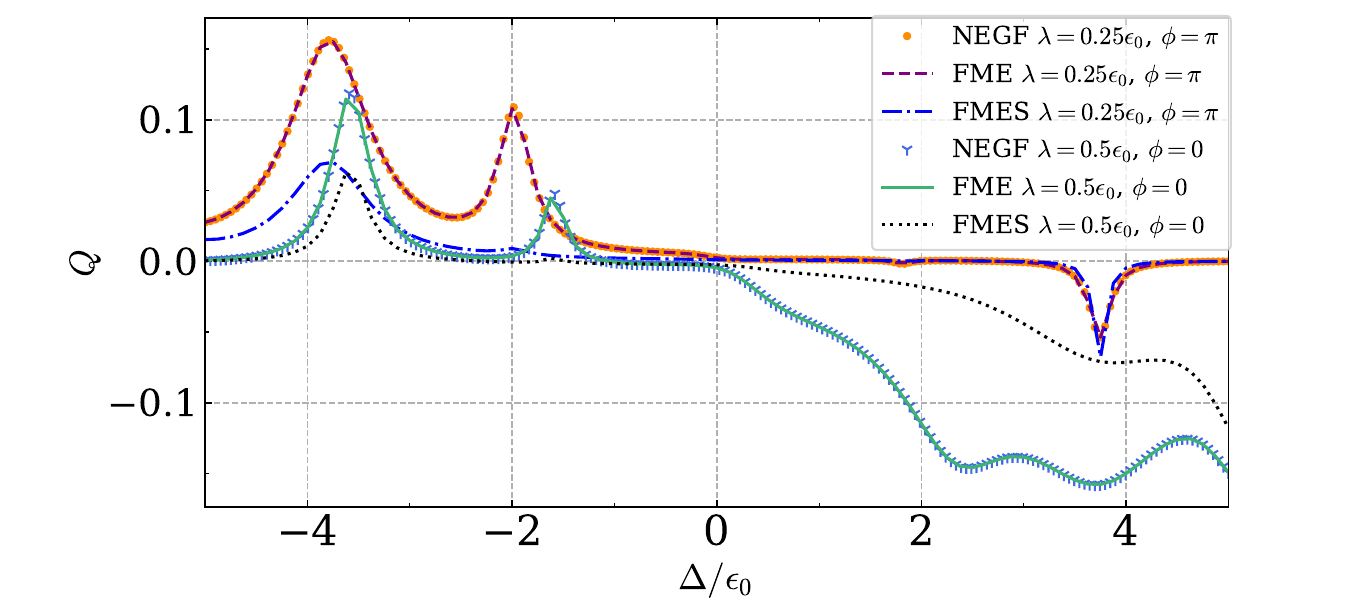} 
\caption{Total pumped charge per period for the same parameters as in Fig.~\ref{Fig:Density} comparing the method of FME used in sec.~\ref{sec: high freq}, TNEGF and FMES}
\label{Fig:Benchmark}
\end{figure}

In sec.~\ref{sec: high freq} we studied our pump model in the high frequency regime by applying a RC mapping for each reservoir and using a Floquet master equation (FME). Figure \ref{Fig:Benchmark} shows a comparison between the results obtained using that method and results obtained using the framework of time-dependent non-equilibrium Green's functions (TNEGFs) \cite{Jauho1994,Aoki2014,Kohler2005,Economou2006}. The results are indistinguishable, verifying the adequacy of the proposed method.\\

The basic idea of the TNEGF approach consists in splitting the Hamiltonian into driven and undriven parts.  Then, one uses as a starting point the Green's function in the long-time limit of the undriven part and applies Floquet theory to obtain the Green's function of the full model in frequency space. The transformations needed to arrive from the time-domain to frequency space (Floquet representation) and vice-versa are nicely illustrated in~\cite{Aoki2014}. The knowledge of the Floquet Green's function then enables us to calculate particle currents just as presented in \cite{Jauho1994}. \\

The time-independent Green's function in the long-time limit can easily be obtained for any non-interacting tight-binding system assuming reservoirs have a semi-circle shaped SD~\cite{Economou2006}.  In order to compare the TQD system studied in sec.~\ref{sec: high freq},  to a TQD coupled to two reservoirs with a semi-circle shaped SD described by the Green's function method, one has to set the radii of the semi-circles large enough to approximate a flat band limit.  \\

The method of FME presented grants access to fluctuations directly (see sec. IV) and can handle interactions within the supersystem without the need of any further approximations. Characteristics like that can become challenging to handle with the TNEGF method.\\

Figure \ref{Fig:Benchmark} also shows results obtained from the FME with the secular approximation performed (FMES). In general, we see that the secular Floquet master equation for the TQD fails to predict the current accurately. The failure of the secular approximation for the master equation of the supersystem regarding RC mappins has been observed before in both driven \cite{Restrepo2018} and undriven \cite{Nazir_Gernot2018, Strasberg2016, Strasberg2018} systems.

\section{Parallel and Series QD}\label{App:Series}
\subsection{Mapping}

The Hamiltonian of the driven TQD, equation~(\ref{eq:HTQD}) can be written as
\begin{equation}
H_{TQD}(t) = \mathbf{d^{\dagger}}\,  \mathbf{H}_{TQD}(t) \,  \mathbf{d},
\end{equation}
where $\mathbf{H}_{TQD}(t)$ is a $3\times 3$ matrix and $\mathbf{d}$ a 3 dimensional vector given by
\begin{align} \label{eq:TQD_matrix}
  \mathbf{H}_{TQD}(t) &=  
\begin{pmatrix} 
\epsilon_L & -\lambda_L(t) & 0 \\ 
-\lambda_L(t) & \epsilon(t) & -\lambda_R(t) \\
0 &  -\lambda_R(t)  & \epsilon_R\\
\end{pmatrix} 
, \\
\mathbf{d} &= 
\begin{pmatrix} 
d_L\\ 
d \\
d_R\\
\end{pmatrix} .
\end{align}
The matrix $  \mathbf{H}_{TQD}(t)$ can now be diagonalized such that
\begin{equation} 
 \mathbf{H}_{TQD}(t)  =  \mathbf{T}^{\dagger}(t)  \mathbf{H}_{P}(t) \mathbf{T}(t),
\end{equation}
where $\mathbf{H}_{P}(t)$ is a diagonal matrix with entries labeled as $\varepsilon_{u}(t),$ $\varepsilon_{c}(t)$ and $\varepsilon_{d}(t)$ and $\mathbf{T}(t)$ is a time-dependent diagonalization transformation. Defining new creation and annihilation operators
\begin{equation} \label{eq:c_def}
 \mathbf{c}(t)  =\mathbf{T}(t) \mathbf{d}, \qquad 
\mathbf{c}(t) = 
\begin{pmatrix} 
c_{u}(t)\\ 
c_{c}(t) \\
c_{d}(t)\\
\end{pmatrix},
\end{equation}
we have
\begin{equation} 
H_{TQD}(t)  =    \sum_{i=\left\lbrace u,c,d  \right\rbrace } \varepsilon_i(t) c^{\dagger}_i(t) c_i(t). 
\end{equation}
It might be tempting to think that the of three parallel QDs constitute three independent transport channels where the net matter current is given by the sum of the current through the three channels. Nevertheless, due to the time dependence in transformation $\mathbf{T}(t)$, this is in general not the case. We look at operator $c_i(t)$ in the interaction picture. It is given by
\begin{equation} \label{eq:C_interaction}
\tilde{c}_i(t) =  \mathcal{U}^{\dagger}(t) c_i(t) \mathcal{U} (t),
\end{equation}
with operator $\mathcal{U} (t)$ satisfying the differential equation $i \partial_t \mathcal{U}(t) = H_{TQD}(t) \mathcal{U}(t) $. Taking the time derivative of equation~(\ref{eq:C_interaction}) we have
\begin{equation}
\begin{aligned}
\dot{\tilde{c}}_i(t) &=   i \, \mathcal{U}^{\dagger}(t)  \left[ H_{TQD}(t), c_i(t)  \right] \mathcal{U}(t)   + \mathcal{U}^{\dagger}(t) \, \dot{c}_i(t) \, \mathcal{U} (t)\\
&= - i \varepsilon_i(t) \tilde{c}_i(t) + \mathcal{U}^{\dagger}(t)  \sum_{i,j }\dot{T}_{ij}(t) d_j  \, \mathcal{U} (t) \\
& =  - i \varepsilon_i(t) \tilde{c}_i(t) + \sum_{j,k} \dot{T}_{ij}(t) T^*_{kj}(t) \tilde{c}_k(t),
\end{aligned}
\end{equation}
where a dot indicates a derivative with respect to time $\dot{h}(t) = \frac{d}{dt} h(t)$.   For the case of an un-driven TQD, matrix $\mathbf{H}_{TQD}$ and transformation $\mathbf{T}$ will be time independent such that $\dot{T}_{ij}(t) = 0$ and therefore for the creation operator $c_i(t)$ we obtain
\begin{equation} 
\dot{\tilde{c}}_i(t) = - i \varepsilon_i \tilde{c}_i(t),
\end{equation}
which is the Heisenberg equation of motion corresponding to the Hamiltonian of a QD with energy $\varepsilon_i$.

\subsection{Rate equation for each channel}

In the low frequency regime our model maps to three independent QDs. Each one of the obeys an equation of the form
\begin{widetext}
\begin{equation} \label{eq:Rate_eq}
 \begin{aligned}
\partial_t \begin{pmatrix} 
P_0(\xi,t) \\ 
P_1(\xi,t)
\end{pmatrix} = - L(\xi,t) \begin{pmatrix} 
P_0(\xi,t) \\ 
P_1(\xi,t)
\end{pmatrix}, \quad
L(\xi,t) = 
\begin{pmatrix} 
-f(t) [ \Gamma_{i,R}(t) + \Gamma_{i,L}(t) ] & [1-f(t)][ \Gamma_{i,R}(t) + e^{-\ii\xi} \, \Gamma_{i,L}(t)] \\ 
f(t) [  \Gamma_{i,R}(t) + e^{\ii\xi} \,\Gamma_{i,L}(t) ] & -[1-f(t)][\Gamma_{i,R}(t) + \Gamma_{i,L}(t)]
\end{pmatrix}, 
\end{aligned}
\end{equation}
\end{widetext}
with $f(t)= f(\varepsilon(t))$ a time dependent Fermi distribution, $P_0(0,t) = 1 - n(t)$, $P_1(0,t) = n(t)$ and $n(t)$ the occupation of the dot. Such equation is sometimes referred to as an adiabatic equation since it follows from assuming that at every instant in time the QD is on its instantaneous energy eigenstate. Using equations (\ref{eq:Cumulants}) and (\ref{eq:Cumulants_X}) we obtain
\begin{equation} \label{eq:Cumulant_Rate_Eq}
 \begin{aligned}
I(t) &= \Gamma_{i,L}(t) \left[  f(t) - n(t)   \right], \\
S(t) &=  \Gamma_{i,L}(t) \left[ f(t) + n(t) - 2 f(t) n(t) \right. \\
& \quad  \left. + 2 f(t) X_0(t) - 2 (1-f(t)) X_1(t)    \right] 
\end{aligned}
\end{equation}
with the auxiliary equations
\begin{equation} 
 \begin{aligned}
\dot{X}_0 (t)  &= -f(t) \Gamma_{i,+}(t) X_0(t)  +  \left[ 1 - f(t) \right]    \Gamma_{i,+}(t) X_1(t)  \\ &\quad - I(t)(1 - n(t)) - \left[ 1 - f(t) \right]  \Gamma_{i,L}(t) n(t)  , \\
\dot{X}_1 (t)  &= f(t) \Gamma_{i,+}(t) X_0(t)  - \left[ 1 - f(t) \right]  \Gamma_{i,+}(t) X_1(t) \\ &\quad+ \Gamma_{i,L}(t)f(t)\left[ 1 - n(t) \right]   - I(t)n(t), 
\end{aligned}
\end{equation}
where $\Gamma_{i,+}(t) = \Gamma_{i,R}(t) + \Gamma_{i,L}(t)$. To obtain the rates $\Gamma_{i,\nu}$ of each
parallel dot one can invert equation~(\ref{eq:c_def}), rewrite the interaction terms between the TQD and the residual baths in equation~(\ref{eq:H_mapped}) in terms of the operators $c_i$ and $c_i^{\dagger}$, and apply Fermi's golden rule.




\bibliography{Mybib_new}


\end{document}